\documentclass{article}
\usepackage[utf8]{inputenc}
\usepackage{geometry}
\usepackage{auxlib}
\usepackage{authblk}

\usepackage{biblatex}
\usepackage[normalem]{ulem} %for strikethrough of text using \sout{} command
\newcommand{\agr}{\mathsf{agr}}

\newcommand{\sch}{\mathsf{s}}
\newcommand{\st}{\mathsf{st}}

\newcommand{\lt}{\mathsf{left}}
\newcommand{\rt}{\mathsf{right}}

\newcommand{\slot}{\mathsf{st}}

\newcommand{\rank}{\mathsf{rank}}
\newcommand{\cost}{\mathsf{ct}}

\newcommand{\hcI}{\hat{\mathcal{I}}}
\newcommand{\hj}{\hat{j}}

\newcommand{\onet}{\texttt{OneEvent}}

\def\PSBA/{\textsc{PESBA}}

\newcommand{\gp}{\texttt{GoodPosn}}

\title{Public Event Scheduling with Busy Agents\thanks{All authors (ordered alphabetically) have equal contributions and are corresponding authors. 
}}

\author[1]{Bo Li}
\author[2]{Lijun Li}
\author[2]{Minming Li}
\author[3]{Ruilong Zhang}

\affil[1]{\small Department of Computing, 
The Hong Kong Polytechnic University}
\affil[2]{\small Department of Computer Science, City University of Hong Kong}
\affil[3]{\small Department of Computer Science and Engineering, University at Buffalo}

\affil[ ]{\texttt{comp-bo.li@polyu.edu.hk, lijunli3-c@my.cityu.edu.hk, minming.li@cityu.edu.hk, ruilongzhang.cn@gmail.com}}

\date{}

\begin{document}

\maketitle

\begin{abstract}
We study a public event scheduling problem, where multiple public events are scheduled to coordinate the availability of multiple agents.
The availability of each agent is determined by solving a separate flexible interval job scheduling problem, where the jobs are required to be preemptively processed.
The agents want to attend as many events as possible, and their agreements are considered to be the total length of time during which they can attend these events.
The goal is to find a schedule for events as well as the job schedule for each agent such that the total agreement is maximized.

We first show that the problem is $\NP$-hard, and then prove that a simple greedy algorithm achieves $\frac{1}{2}$-approximation when the whole timeline is polynomially bounded.
Our method also implies a $(1-\frac{1}{e})$-approximate algorithm for this case.
Subsequently, for the general timeline case, we present an algorithmic framework that extends a $\frac{1}{\alpha}$-approximate algorithm for the one-event instance to the general case that achieves $\frac{1}{\alpha+1}$-approximation.
Finally, we give a polynomial time algorithm that solves the one-event instance, and this implies a $\frac{1}{2}$-approximate algorithm for the general case.

\end{abstract}

\newpage
\section{Introduction}

Artificial intelligence algorithms are widely used in many societal settings, such as scheduling public activities and assisting with personal work schedules, in which finding a suitable schedule shall lead to good social welfare.
To motivate our study, consider the pubic event scheduling process at a university. 
Suppose the student/staff development office wants to hold some public activities for students/staffs, e.g., networking events, new student/staff orientation, etc.
These events have fixed duration, while the target students/staffs are busy, and they have their own tasks that must be done but with some degree of time flexibility, e.g., homework, lectures, research meetings, submission deadlines, etc.
For example, there are two events, both lasting for two hours. 
Some of the students/staffs must spend two hours on their own tasks between 10:00 and 13:00, and other students/staffs must spend three hours between 14:00 and 18:00. 
The whole schedulable time span for events is 10:00-18:00. 
These events are important to students/staffs, and it is crucial to schedule these public events at suitable times so that the target students/staffs can participate as long as possible.
The {\em agreement} of a student/staff to an event schedule is considered as the total time that she can attend the events.
Hence, the organizer's task is to make the event schedule admit a maximum agreement.
This shall lead to good social welfare, motivating our work.

The above public event scheduling problem aims to find feasible job and event schedules such that social welfare is maximized.
Hence, it falls under the umbrella of both job scheduling and computational social choices.
The intersection of these two fields has been extensively studied in various literature, such as selfish load balancing~\cite{DBLP:conf/wine/BiloMMV20,DBLP:journals/teco/VinciBMM22}, the Santa clause problem~\cite{DBLP:conf/icalp/BamasGR21,DBLP:conf/stoc/BansalS06,DBLP:conf/nips/SpringerHPK22}, fair division with scheduling constraints~\cite{DBLP:conf/nips/LiLZ21,DBLP:conf/nips/LiWZ23,DBLP:conf/icml/0002B023},  etc.
% Following the conventions of the scheduling community, we model the personal tasks of a student or staff as a job that is associated with a release time, processing time, and deadline; all of these jobs need to be finished by their deadlines, where we allow preemption; the processing of a job can be interrupted by the other jobs' processing.
% By following the conventions of computational social welfare, we model a student or staff as an agent, and the objective is considered to be social welfare maximization, i.e., we aim to maximize the total agreement. 

From the scheduling perspective, our problem shares some similarities with a set of scheduling problems.
Intuitively, the optimal solution to our problem tends to gather the schedule of agents' jobs, which leads to a long consecutive idle time during which events can be scheduled.
The optimal solution for a batch of scheduling problems also shares a similar behavior, such as gap scheduling~\cite{DBLP:conf/soda/AntoniadisGKK20,DBLP:conf/soda/Baptiste06}, active time slots minimization~\cite{DBLP:journals/algorithmica/ChangGK14,DBLP:conf/birthday/ChauL20,DBLP:conf/spaa/KumarK18}, calibration scheduling~\cite{DBLP:conf/spaa/BenderBLMP13,DBLP:journals/tcs/ChauLWZZ20,DBLP:conf/spaa/ChenLL019}, etc.
However, our work investigates a distinct objective compared to these classical scheduling problems and thus leads to completely different techniques.

From the perspective of computational social choice, our problem shares some similarities with the recently proposed cake-sharing model~\cite{DBLP:conf/aaai/BeiLS22}.
We can consider the whole timeline as a cake whose range is from $[0,1]$, and then normalize all agents' tasks to ensure that jobs' release time and deadline are in $[0,1]$.
Furthermore, we assume the jobs' processing time is equal to the length of its interval, where we call a job {\em rigid} (otherwise, it is called {\em flexible}).
An agent's task can be interpreted as a preference for the cake, i.e., if an agent has a task $[a,b]$, then the agent does not prefer this part of the cake.
Our goal is to pick $m$ pieces of cakes such that the total agreement of the picked cakes is maximized.
The above scenario also motivates our work, and it matches the setting of cake-sharing~\cite{DBLP:conf/aaai/BeiLS22} in the general sense.
Compared to cake sharing, the first main difference is that we allow jobs to be flexible. 
The second main difference is that we focus on the social welfare maximization problem while~\cite{DBLP:conf/aaai/BeiLS22} aims to design truthful mechanisms.

\subsection{Our Contributions}

We consider the problem of Public Event Scheduling with Busy Agents (\PSBA/).
We distinguish several cases according to the event set or the length of the whole timeline.
For each case, we either give an approximate (or exact) algorithm running in polynomial time or show the problem is $\NP$-hard.

\paragraph{Main Result 1 (\cref{thm:hardness}, \ref{thm:one-event}).}
The problem \PSBA/ is $\NP$-hard even in the case where (\rom{1}) there is only one agent; (\rom{2}) this agent has only two rigid jobs.
Moreover, the problem \PSBA/ is in $\PP$ when there is only one event that is required to be scheduled.

Our first result focuses on the computation complexity of the problem (\cref{sec:hardness} and \cref{sec:poly:one-event}).
We first show that \PSBA/ is $\NP$-hard even if there is only one agent and the agent has only two rigid jobs.
The hardness result is built on the classical partition problem.
In \cref{sec:poly:one-event}, we show \PSBA/ is in $\PP$ when there is only one event that needs to be scheduled.
This result shall be used to get an approximation for general \PSBA/ instances.
Our algorithm for one-event instance is involved, and its basic idea is to draw the plot for the agreement function when the event moves over the timeline.
To this end, we prove that the agreement function is piecewise linear and has a polynomial number of turning points.

\paragraph{Main Result 2 (\cref{thm:pseudo}).}
When the whole timeline is polynomially bounded, a natural greedy algorithm running in polynomial time achieves $\frac{1}{2}$-approximation for \PSBA/.
Moreover, \PSBA/ admits a $(1-\frac{1}{e})$-approximate algorithm running in polynomial time.

For a better understanding of our method, let us first focus on the case where the whole timeline is polynomially bounded (\cref{sec:pseudo}).
Our main algorithm is an intuitive greedy algorithm which can be viewed as a simple voting process.
Namely, in each round of our algorithm, for each unscheduled event, we ask agents to vote on the position for this event, and each agent votes for those time slots that maximize her agreement.
The algorithm then picks one position that receives the most votes for this event.
Finally, the algorithm schedules the event that maximizes the total agreement among all unscheduled events in this round.
By proving (\rom{1}) a new result for a variant of the submodular maximization (\rom{2}) the agreement function is submodular, we show that the greedy algorithm is $\frac{1}{2}$-approximate.
This intuitive algorithm enables us to extend it to arbitrary timeline cases.
In fact, we show a stronger result that the agreement function is a rank function of some matroid, which is also crucial for the arbitrary timeline case.
By using the more involved algorithms for submodular optimization with matroid constraints, our method also implies a better $(1-\frac{1}{e})$-approximate algorithm, but it is hard to extend to arbitrary timeline cases.

\paragraph{Main Result 3 (\cref{thm:general-T}).}
When the whole timeline is arbitrary, we present an algorithmic framework that extends a $\frac{1}{\alpha}$-approximate algorithm for \PSBA/ with one event to the multiple events case achieving $\frac{1}{\alpha+1}$-approximation. We design an optimal algorithm for the one-event instance and hence obtain a $\frac{1}{2}$-approximate algorithm for the arbitrary timeline case.

Our second result focuses on the general \PSBA/ instance (\cref{sec:poly}).
We propose an algorithmic framework for the general case.
Given any $\frac{1}{\alpha}$-approximate algorithm for the one-event instance, our framework extends the algorithm to multiple events case in polynomial time that achieves $\frac{1}{\alpha+1}$-approximation.
Our algorithmic framework is rooted in the greedy algorithm for the polynomially bounded timeline case.
In each iteration of the greedy algorithm, it enumerates each time slot for searching the best position of each unscheduled event. This only works for the previous case since the whole timeline is polynomially bounded.
We design a new oracle based on the given one-event approximation algorithm that is able to return a ``good'' position for each unscheduled event without enumerating the whole timeline.
How good the position is depends on the approximation ratio of the given one-event algorithm.
By using the optimal algorithm for the one-event instance proposed in \cref{sec:poly:one-event}, we directly obtain a $\frac{1}{2}$-approximate algorithm for general \PSBA/ instances, which matches the approximation ratio for restricted timeline case.

\subsection{Other Related Works}

\paragraph{Fair Allocation with Public Goods.}
Our problem shares some similarities with fair allocation of public goods.
\cite{DBLP:conf/sigecom/ConitzerF017} first uses ``public'' to distinguish the previous ``private'' goods, and subsequently, there is a batch of follow-up works~\cite{DBLP:conf/sigecom/FainM018,DBLP:conf/aaai/FluschnikSTW19,DBLP:conf/fsttcs/GargKM21}.
In the public goods setting, all agents obtain utilities when some goods are selected, which is similar to the agreement in our problem.
However, the techniques are different since we consider a different objective.
% i.e., we focus on a solution that maximizes the total social welfare, while they focus on a solution that is fair among all agents.

\paragraph{Other Related Scheduling Problems.}
When the preemption is not allowed, it is $\NP$-complete to determine whether the given job set admits a feasible schedule~\cite{DBLP:books/fm/GareyJ79}.
When all jobs are rigid or unit, the problem is equivalent to determining the size of the maximum independent set on interval graphs or a maximum matching. Hence it is in $\PP$~\cite{DBLP:books/daglib/0090562}.
% When all jobs has a unit processing time, the problem is equivalent to a maximum matching problem; so, it is also in $\PP$~\cite{DBLP:books/daglib/0090562}.
When preemption is allowed, a simple greedy algorithm (earliest deadline first) can be used to decide instances' feasibility~\cite{DBLP:journals/corr/abs-2001-06005}.

\section{Preliminaries}\label{model}

We consider the problem of Public Event Scheduling with Busy Agents (\PSBA/).
An instance of \PSBA/ consists of an agent set $A:=\set{1,\ldots,n}$ and a public event set $E:=\set{e_1,\ldots,e_m}$.
We consider discrete time, and for $t\in\N_{\geq 1}$, let $[t,t+1)$ denote the $t$-th time slot.
$T$ is a time slot set consisting of all time slots where events and jobs can be scheduled.
Each agent $i$ has a set of jobs denoted by $J_i$.
Each job $j$ in $J_i$ has a release time $r_j\in \N_{\geq 1}$, a deadline $d_j\in\N_{\geq 1}$, and a processing time $p_j\in \N_{\geq 1}$ such that $p_j\leq d_j-r_j+1$.
A job $j$ is called a {\em rigid} job if $p_j=d_j-r_j+1$; it is called a {\em unit} job if $p_j=1$.
For each job $j$, the interval $[r_j,d_j]$ is called a {\em job interval}, which can also be viewed as a set of consecutive time slots, i.e., $\set{r_j,r_j+1,\ldots,d_j}$.
The event set $E$ consists of $m$ public events, and each event $e$ has a length $l(e)\in\N_{\geq 1}$.
When an event $e$ is scheduled at the $t$-th time slot, this event occupies $\set{t,t+1,\ldots,t+l(e)-1}$ time slots to be held.
We use a tuple $(e,t)$ to denote that event $e$ is scheduled at the $t$-th time slot.
Let $\cL$ be a set of all possible tuples, i.e., $\cL:=\set{(e,t)}_{e\in E,t\in T}$.
Let $\cL(e)$ be the set of all possible schedules of event $e$, i.e., $\cL(e):=\set{(e',t)\in \cL\mid e'=e}$.
% \lijun{To see whether $\cL(e)$ is used elsewhere}

A solution consists of both the event schedule and the job schedule for each agent.
The event schedule $\cS$ is considered as a subset of $\cL$ such that every event appears in exactly one tuple in $\cS$.
This ensures that (\rom{1}) every event $e\in E$ is scheduled at some time slot; (\rom{2}) no event is scheduled more than two time slots.
An event schedule $\cS$ is said to be a partial event schedule if every event appears in at most one tuple; otherwise, it is called a complete event schedule.
The job schedule consists of $n$ subschedules, each of which corresponds to a job schedule for each agent.
We consider the preemptive schedule, and the job schedule for each $J_i,i\in A$ is required to be feasible, i.e., (\rom{1}) each job $j$ in $J_i$ is assigned $p_j$ time slots in its job interval; (\rom{2}) for each $t\in T$, at most one job from $J_i$ is processed at $t$.
It is ensured that for each $i\in A$, the input job set $J_i$ is feasible.

Each event is considered to be a good activity, and agents are willing to attend the events.
A time slot $t$ is said to be an {\em agreement} time slot for agent $i$ if (\rom{1}) $t$ is occupied by some scheduled event; (\rom{2}) no job from $J_i$ is scheduled at $t$.
Given any event schedule and a job schedule for $J_i$, agent $i$'s {\em agreement} is defined as the total number of agreement time slots.
We allow event schedules to overlap since we can always eliminate the overlap by adjusting events' schedules without decreasing the total agreement (see \cref{shiftevent}).
Our goal is to find a job schedule for each $J_i$ as well as a complete event schedule such that the total agents' agreement is maximized.

\begin{figure}[htb]
    \centering
    \includegraphics[width=6cm]{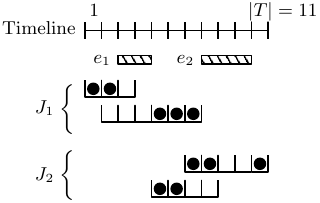}
    \caption{Illustration of the model of PSBA problem. There are two public events $E=\set{e_1,e_2}$ with length $l(e_1)=2$ and $l(e_2)=3$ represented by shaded rectangles, and two agents $A=\set{1,2}$. 
    Agent $1$'s job set is $J_1$ containing two jobs $[1,3]$ with processing time $2$ and $[2,7]$ with processing time $3$. 
    Agent $2$'s job set is $J_2$ containing two jobs  $[7,11]$ with processing time $3$ and $[5,8]$ with processing time $2$.
    The time span is $T=[1,11]$ containing $11$ time slots.
    The figure shows one optimal event schedule $\cS=\set{(e_1,3),(e_2,8)}$ maximizing $\sum_{i\in A}\agr_i(\cS)$ and a corresponding optimal job schedule where the black disk indicates job's processing.
    Agent $1$ can attend the whole course of $e_1$ and $e_2$ under schedule $\cS$ and thus get agreement $5$; agent $2$ can attend whole $e_1$ but can only attend $e_2$ for $2$ times and thus get agreement $4$ from $\cS$.}
    \label{fig:model}
\end{figure}

We use function $\agr_i:2^{\cL}\to\N_{\geq 0}$ to represent the maximum agreement that is produced by an event schedule $\cS$ for agent $i$.
Given any partial event schedule $\cS$ and an agent $i$, in \cref{sec:pseudo:agr}, we show that the value of $\agr_i(\cS)$ can be computed in polynomial time.
Such an algorithm is used to compute the optimal job schedule, and this allows us to focus only on finding an event schedule.
Based on the agreement function, our objective is equivalent to finding a complete event schedule $\cS$ such that $\sum_{i\in A}\agr_i(\cS)$ is maximized.
An example can be found in \cref{fig:model}.
In the remainder of this paper, we refer to $\cS$ as a {\em partial} solution if $\cS$ is a partial event schedule; otherwise, it is a {\em complete} solution.

\begin{figure}[H]
    \centering
    \includegraphics[width=6cm]{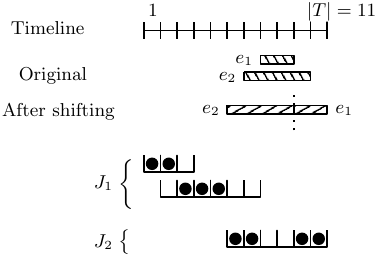}
    \caption{Originally $e_1$ has overlap with $e_2$ which starts before it and thus we delay $e_1$ to $[11,13)$ and then shift $e_1$ and $e_2$ together to $[6,12)$. The black disks indicate jobs' processing. We can see the shift increases the number of agreement time slots of $J_1$'s schedule from $4$ to $6$ while keeping the number of agreement time slots of $J_2$'s schedule unchanged.}
    \label{fig:shiftevt}
\end{figure}

\begin{observation}\label{shiftevent}
Given any solution $\cS$ with total agreement $\sum_{i\in A} \agr_i(\cS)$, there exists a solution $\cS'$ such that (i) $\sum_{i\in A} \agr_i(\cS')\geq \sum_{i\in A} \agr_i(\cS)$, (ii) the event schedule in $\cS'$ has no overlap.
\end{observation}
\begin{proof}
    Given any solution $\cS$ with total agreement $\sum_{i\in A} \agr_i(\cS)$, we shift events in $\cS$ to eliminate overlap with keeping the job schedule unchanged. 
    First, we assume $|T|$ is infinite. Order events ascendingly according to their starting time (break ties arbitrarily) and then repeat this process: for the first event $e$, which overlaps with one event $e'$ whose starting time is before $e$, delay $e$ until $e$ has no overlap with $e'$ (after the delay $e$ will start from the ending time of $e'$). After this, if there are events scheduled outside of the actual time span $T$, we take all events as a whole and shift them together towards an earlier time until they are all scheduled within $T$. Let $\cS'$ be the new event schedule after such shift. 
    
    The agreement of $\cS'$ won't be smaller than that of $\cS$, because considering an arbitrary agreement time slot of any agent, this time slot is still an agreement time slot after shift. An example can be found in \cref{fig:shiftevt}.
\end{proof}

\section{Hardness}
\label{sec:hardness}

In this section, we show that \PSBA/ is $\NP$-hard even in the case where (\rom{1}) there is only one agent; (\rom{2}) all jobs are rigid; see \cref{thm:hardness}.

\begin{theorem}\label{thm:hardness}
Given an instance of \PSBA/ such that (\rom{1}) $\abs{A}=1$; (\rom{2}) $\abs{J_1}=2$ and all jobs are rigid, it is $\NP$-complete to determine whether the maximum agreement is $Q$, where $Q$ is some given parameter.
\end{theorem}

\begin{proof}
We reduce the Partition problem to a special case of our problem where all jobs are rigid. 
Given any instance of the partition problem consists of a set of integers $S=\{s_1,s_2,\ldots,s_m\}$ with $Q=\frac{1}{2} \sum_{i=1}^m s_i$.
We construct a rigid instance $(E,A,J,T)$ of \PSBA/ as follows: for each $s_i$, there is an event $e_i \in E$ with length $l(e_i)=s_i$. 
There is only one agent who has $2$ rigid jobs $j_1$ and $j_2$; the release time of $j_1$ is $1$ and the deadline is $Q$; the release time of $j_2$ is $2Q$ and the deadline is $3Q-1$. 
The total time span is $\abs{T}=4Q$. 
We ask whether there is an event schedule with the largest agreement being $2Q$, that is, the agent's job schedule has no overlap with the event schedule. 

Given a partition solution $S^1$ and $S^2$, schedule the event with length $s_i \in S^1$ one by one dis-jointly in the first ``job-free area'' $(Q,2Q)$; and schedule the event with length $s_i\in S^2$ one by one dis-jointly in the second ``job-free area'' $(3Q,4Q]$. This schedule's agreement is $2Q$ since no events have overlap with jobs.

Given an event schedule with $2Q$ agreement of the \PSBA/ problem, then we must have two sets of events scheduled within $(Q,2Q]$ and $(3Q,4Q]$ respectively each of which has total event length equal to $Q$. For $e_i$ with $\sch(e_i) \subseteq (Q,2Q]$, we let $s_i \in S^1$, and for $e_i$ with $\sch(e_i) \subseteq (3Q,4Q]$, we let $s_i \in S^2$. Then the sum of elements in $S^1$ is equal to that in $S^2$ which is $Q$, meaning an equal partition.
\end{proof}
\section{Algorithms for Polynomial T}
\label{sec:pseudo}

In this section, for a better understanding of our method, we focus on the case where the whole timeline is polynomially bounded; the assumption shall be removed in \cref{sec:poly}.
We show that a natural greedy algorithm achieves $\frac{1}{2}$-approximation.
In \cref{sec:poly}, we shall extend the $\frac{1}{2}$-approximation algorithm to the arbitrary $\abs{T}$ case with the approximation ratio being preserved.
Our method also implies a slightly stronger algorithm that achieves $(1-\frac{1}{e})$-approximation in the polynomial bounded $\abs{T}$ case.
Such a stronger result uses the submodular maximization subject to a matroid constraint~\cite{DBLP:journals/siamcomp/CalinescuCPV11} as a black box, and thus, it is hard to be extended to the arbitrary $\abs{T}$ case.
Thus, we focus on the greedy algorithm, and we state how to get a $(1-\frac{1}{e})$-approximation in the proof of \cref{thm:pseudo}.

\begin{theorem}
Given any instance of \PSBA/ with polynomially bounded $\abs{T}$, there exists a greedy algorithm running in polynomial time that achieves $\frac{1}{2}$-approximation.
The approximation ratio can be further improved to $(1-\frac{1}{e})$.
\label{thm:pseudo}
\end{theorem}

\paragraph{Algorithmic Framework.}
For each event $e\in E$ and a time slot $t\in T$, the tuple $(e,t)$ represents that the event $e$ is scheduled at the $t$-th time slot.
Since $\abs{T}$ is assumed to be polynomially bounded, we can find the optimal position for event $e$ by enumerating all possible positions. 
% Let $\cL$ be the set of all possible tuples, i.e., $\cL$ includes all possible positions for each event.
% We use $\cL(e)\subseteq \cL$ to represent all possible schedule times of event $e$, i.e., $\cL(e):\set{(e',t)\in \cL\mid e'=e}$.
% Since $\abs{T}$ is assumed to be polynomially bounded, the size of each $\cL(e)=\abs{T}$ is polynomial; this implies that the size of $\cL$ is also polynomial in the input size.
Our algorithm runs in rounds and schedules exactly one event in each round until all events are assigned to a starting position.
In each round, the algorithm checks the unscheduled events one by one.
For each unscheduled event $e'$, every agent assigns some ``votes'' to every possible position of $e'$.
Then, our algorithm chooses the schedule $(e',t')$ that receives the most votes (break ties arbitrarily), which gives a scheduled position $t'$ for event $e'$.
For each unscheduled event, we have a position in this round, and then, the algorithm just picks an event and its position $(e^*,t^*)$ with the maximum votes among all unscheduled events.
Intuitively, the votes of some schedule $(e,t)$ stands for the agent's preference for this schedule. 
Our algorithm shall select the schedule that is liked by the most agents and schedule the corresponding event at the corresponding time slot.
% In other words, our algorithm repeats the following steps until all events are scheduled:

The first step of the algorithm is to set up an appropriate preference rule such that each agent assigns suitable votes to each schedule.
Our idea is intuitive, i.e., agent $i$'s preference for some schedule $(e,t)$ is just the maximum incremental agreement that agent $i$ can gain by scheduling $e$ at $t$, i.e., $\agr_i(\cS\cup\set{(e,t)})-\agr_i(\cS)$ where $\cS$ is the current solution.

\subsection{The Complete Algorithm}

This section presents the formal description of the algorithm.
Given the partial solution $\cS$, we define $F$ to be the total agreement produced by $\cS$ for notation convenience, i.e., $F(\cS):=\sum_{i\in A}\agr_i(\cS)$.
For an unscheduled event $e$ and any schedule $(e,t)$, the total preference for the schedule $(e,t)$ is $F(\cS\cup\set{(e,t)})-F(\cS)$.
Following the basic idea stated above, the algorithm picks a schedule with the maximum total preference in each round until all events are scheduled.

\begin{algorithm}[H]
\caption{The Complete Algorithm for \PSBA/}
\label{alg:pseudo}
\begin{algorithmic}[1]
\Require A \PSBA/ instance with polynomially bounded $\abs{T}$.
\Ensure A schedule $\cS$ of all events in $E$.
\State $\cS\leftarrow\emptyset$.
\While{there exists an unscheduled event}
\State $\cH\leftarrow\emptyset$.
\For{each unscheduled event $e'$}
\State $t' \leftarrow\argmax_{t\in T} F(\cS\cup\set{(e',t)})-F(\cS)$.
\label{line:pseudo:max}
\State $\cH\leftarrow \cH\cup \set{(e',t')}$.
\EndFor
\State $(e^*,t^*)\leftarrow \argmax_{(e,t)\in\cH} F(\cS\cup\set{(e,t)})-F(\cS)$.
\State $\cS\leftarrow \cS\cup\set{(e^*,t^*)}$.
\EndWhile
\State \Return $\cS$.
\end{algorithmic}
\end{algorithm}

The analysis of \cref{alg:pseudo} is built on a variant of a classical submodular maximization problem, i.e., the selected elements are additionally required to satisfy the group constraints, and an imperfect oracle exists.
The formal definition is shown in \cref{def:pseudo:submodular-maximization}.

\iffalse
The analysis of \cref{alg:pseudo} is built on the variant of the classical submodular maximization problem, i.e., the selected elements are additionally required to satisfy the group constraints, and an imperfect oracle exists.
As one may observe, the additional group constraint corresponds to the requirement that each event can be assigned at most one starting point in our problem.
The imperfect oracle states that enumerating all elements is not allowed, and to get the element with the maximum marginal value, the algorithm has to ask an imperfect oracle which may only be able to return some non-optimal element; see \cref{def:pseudo:submodular-maximization} for a formal definition.
The imperfect oracle is used to deal with the arbitrary $T$ case in which we are not allowed to enumerate all $t\in T$ in \cref{sec:poly}; it can be ignored in this section.
% Based on the above theorem, we show that the agreement function $\agr_i$ is submodular, so greedily selecting the element that maximizes the marginal value achieves $\frac{1}{2}$-approximation by observing the role of the preference function is equivalent to computing the marginal value.

Before introducing the analysis framework, we first present the definition of the variant of the classical submodular maximization problem (\cref{def:pseudo:submodular-maximization}), which will be used later to show the approximation ratio of \cref{alg:pseudo}.
\fi

\begin{definition}[Submodular Maximization with Group Constraints and Imperfect Oracle (SMGC)]

Given some parameter $n\in \N_{\geq 1}$ and there is a ground element set $U$ with $\abs{U}=2^{\poly(n)}$.
There is a monotone and submodular function $f:2^U\to\R_{\geq 0}$ defined over $U$.
The ground element $U$ is partitioned into $\ell$ groups: $G_1,\ldots,G_{\ell}$ with $\ell=\poly(n)$.
There is a polynomial $\alpha$-approximate oracle $g(\cdot,\cdot)$ that takes two parameters as the input: a subset $S\subseteq U$ and a group $G_i$, and it returns an element $e^*\in G_i$ in $\poly(n)$ time such that $f(S\cup\set{e^*})-f(S)\geq\frac{1}{\alpha}\max_{e\in G_i}f(S\cup\set{e})-f(S)$, where $\alpha\geq 1$ is some given parameter.
The goal is to pick a subset $S$ of the ground element set such that $f(S)$ is maximized, where $S$ is required to satisfy two conditions: (\rom{1}) $\abs{S}\leq k$ ($\ell$ is guaranteed to be larger than $k$); (\rom{2}) $S$ contains at most one element from $G_i$ for all $i\in[\ell]$.
The running time of the algorithm is required to be $\poly(n)$.
\label{def:pseudo:submodular-maximization}
\end{definition}

\paragraph{SMGC v.s. \PSBA/.}
To see the connection, consider $U$ as all possible schedules $\cL$, $n$ as the input size of \PSBA/, $f(\cdot)$ as the total preference function $F(\cdot)$.
The additional group constraint corresponds to the requirement that each event can be assigned at most one starting point in our problem.
The imperfect oracle states that to get an element with a good marginal value in polynomial time, the algorithm has to ask an imperfect oracle, which may only be able to return some non-optimal elements.
The $\alpha$-approximate oracle is used to deal with general $|T|$ cases, and its role will be clear in \cref{sec:poly}.
In this section, we can assume that we have a $1$-approximate oracle (i.e., $\alpha=1$) since we consider the polynomially bounded $|T|$ case.
This implies that the size of $\cL$ (i.e., the ground element set) is polynomially bounded, so finding an element with the maximum marginal value takes polynomial time.
% The imperfect oracle is used to deal with the arbitrary $T$ case in which enumerating all $t\in T$ is not allowed in \cref{sec:poly}; one can pretend that $\alpha=1$ in this section.

% The approximate oracle $g$ is used to handle the arbitrary $T$ case in which enumerating all $t\in T$ is not allowed in \cref{sec:poly}; so, one may pretend that $\alpha=1$ in this section.
When $\alpha=1$, SMGC is a special case of submodular maximization subject to a matroid constraint~\cite{DBLP:journals/siamcomp/CalinescuCPV11}, which admits a $(1-\frac{1}{e})$-approximation.
However, the algorithm is involved, and it is based on the multilinear extension of a submodular function.
% To this end, one has to use the multilinear extension of a submodular function and a dependent rounding algorithm.
For the matroid constraint, the approximate ratio of the natural greedy algorithm is proved to be $\frac{1}{2}$~\cite{horel2015notes} whose proof is based on both the exchange property of a matroid and the submodularity.
In \cref{sec:pseudo:group}, we show that the ratio only depends on the submodularity for our problem and thus simplifies the proof for the $\alpha=1$ case.

Formally, the analysis mainly consists of three steps.
In the first step, we show that the agreement function can be computed in polynomial time.
In the second step, we prove that the agreement function is submodular.
In the last step, we demonstrate that the natural greedy algorithm achieves $\frac{1}{\alpha+1}$-approximation to SMGC with an $\alpha$-approximate oracle; when $\alpha=1$, it is a $\frac{1}{2}$-approximate algorithm.
The detailed description is as follows.

\begin{enumerate}
    \item We show that given any partial solution $\cS$, the value of $\agr_i(\cS)$ can be computed in polynomial time for any $i\in A$.
    This implies that the cumulative function $F$ in \cref{alg:pseudo} can also be computed in polynomial time.
    This part is deferred to \cref{sec:pseudo:agr}.

    \begin{lemma}
    Given any partial solution $\cS\subseteq\cL$, there is an algorithm running in polynomial time that computes the value of $\agr_i(\cS)$.
    \label{lem:pseudo:agr}
    \end{lemma}

    \item We demonstrate that the agreement function $\agr_i(\cdot)$ is a submodular function. 
    We actually show a stronger result, i.e., the agreement function is a rank function of some matroid which also plays an important role in the arbitrary $T$ case.
    This part is shown in \cref{sec:pseudo:submodular}.

    \begin{lemma}
    Consider any agent $i\in A$, the agreement function $\agr_i:2^{\cL}\to \N_{\geq 0}$  is a submodular function.
    \label{lem:pseudo:submodular}
    \end{lemma}

    \item We prove that greedily picking the element with a good marginal value from unselected groups achieves $\frac{1}{\alpha+1}$-approximation for SMGC.
    This part is in \cref{sec:pseudo:group}.
    
    \begin{lemma}
    The natural greedy algorithm (\cref{alg:pseudo:group}) achieves $\frac{1}{\alpha+1}$-approximation to the problem of submodular maximization with group constraints and an $\alpha$-approximate oracle stated in \cref{def:pseudo:submodular-maximization}.
    \label{lem:pseudo:group}
    \end{lemma}
\end{enumerate}

\begin{proof}[Proof of \cref{thm:pseudo}]

We first show that \cref{alg:pseudo} is a $\frac{1}{2}$-approximate algorithm running in polynomial time.
Firstly, since the function $F$ can be computed in polynomial time by \cref{lem:pseudo:agr} and $T$ is polynomially bounded, \cref{alg:pseudo} runs in polynomial time.
Since line~\ref{line:pseudo:max} can be computed in polynomial time, we have $\alpha=1$ in this case.
By \cref{lem:pseudo:submodular}, we know that $\agr_i(\cdot)$ is a submodular function for each $i\in A$; so, $F(\cdot)=\sum_{i\in A}\agr_i(\cdot)$ is also a submodular function.
\cref{alg:pseudo} just greedily picks an element with the maximum marginal value from each unselected group.
Thus, \cref{alg:pseudo} is a $\frac{1}{2}$-approximate algorithm by \cref{lem:pseudo:group}.

We now describe how to get a $(1-\frac{1}{e})$-approximation.
The approximation is based on the following known result for submodular maximization with a matroid constraint.

\begin{claim}[\cite{DBLP:journals/siamcomp/CalinescuCPV11}]
There is a randomized rounding algorithm that achieves $(1-\frac{1}{e})$-approximation for the problem of submodular maximization with matroid constraints.    
\end{claim}

The ground element set $\cL$ can be partitioned into $\abs{E}$ groups, i.e., $\cL(e):=\set{(e',t)\in\cL\mid e'=e}$.
For each group, the algorithm is required to pick exactly one element.
This matches the definition of the partition matroid.
In fact, any subset $\cL'\subseteq \cL$ that contains exactly one element from each $\cL(e)$ is a feasible solution to our problem.
By \cref{lem:pseudo:submodular}, we know that the function $F$ is a submodular function.
Thus, our problem is equivalent to maximizing the submodular function $F$ subject to a partition matroid.
The $(1-\frac{1}{e})$-approximate algorithm runs in polynomial time since the ground element size $\abs{\cL}$ is polynomially bounded.

\end{proof}

\subsection{Agreement Function Computation}
\label{sec:pseudo:agr}

In this section, we show that the maximum agreement of any partial solution can be computed in polynomial time.
We fix an agent $i\in A$ and analyze her agreement function $\agr_i(\cdot)$.
For notation convenience, we drop the subscript $i$ for the time.
% use $\agr(\cdot)$ and $J$ to denote $\agr_i(\cdot)$ and $J_i$, where $J_i$ is agent $i$'s jobs.

\begin{figure}[H]
    \centering
    \includegraphics[width=10cm]{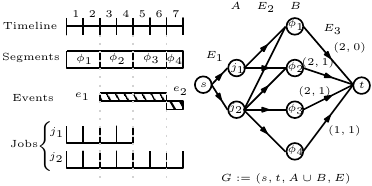}
    \caption{An example for min-cost max-flow construction. The left part of the figure shows the original job set and event schedule with $\Psi=\set{[1,2),[3,4),[4,5),[6,7),[7,8)}$. We cut the whole timeline into a set of segments $\Phi=\set{[1,3),[3,5),[5,7),[7,8)}$ with polynomial size according to $\Psi$. The constructed flow network is shown in the right part of the figure. For each job, we have a vertex in $A$, and for each segment, we have a vertex in $B$. 
    The capacity and the cost of each edge are designed to ensure that a flow assignment corresponds to a job schedule, e.g., $(2,0)$ of the edge in $E_3$ connecting $\phi_1$ and $t$ indicates the capacity of this edge is $2$ and cost is $0$. }
    \label{fig:min-cost-flow}
\end{figure}

For an event or job interval, we use $s(\cdot)$ and $f(\cdot)$ to denote the first and the last time slot occupied by the interval.
Given a partial solution $\cS\subseteq\cL$, let $E'$ be the scheduled event set.
Define $\Psi$ to be the set of $s(j),f(j),s(e),f(e)$ for all $j\in J$ and $e\in E'$ (remove duplicate values).
Note that the size of $\Psi$ is polynomially bounded regardless of the size of the time span $T$.
We sort all time slots in $\Psi$ in increasing order obtaining $\Psi=\set{t_1,\ldots,t_{\kappa}}$.
Then, the time slots in $\Psi$ split the whole time span into several segments, i.e., every adjacent two-time slot in $\Psi$ defines a segment $\phi$. 
See the left part of \cref{fig:min-cost-flow} for an example.
Let $\Phi:=\set{\phi_1,\ldots,\phi_{\ell}}$ be the set of segments.
Note that a segment $\phi$ may contain more than one time slot.
Let $c(\phi)$ be the capacity of the segment $\phi$, i.e., $c(\phi)$ is the number of time slots that are included in $\phi$.
Note that it must be the case that for any segment $\phi\in\Phi$ and any job or event interval $I$, $\phi$ is either completely included in $I$ or $\phi$ has no overlap with $I$.

Our algorithm for computing $\agr(\cdot)$ is built based on the min-cost max-flow algorithm.
In the following, we first introduce the construction of the flow network and then prove the correctness.
Given any partial solution $\cS$ and a job set $J$, we first construct the starting point set $\Psi$ and the segment set $\Phi$.
Then, we construct a network $G:=(s,t,A\cup B, E)$ according to $\Psi$ and $J$.
(\Rom{1}) For each job $j\in J$, we create a vertex $a$ in $A$.
(\Rom{2}) For each segment $\phi\in\Phi$, we have one vertex $b$ in $B$.
(\Rom{3}) There is a directed edge $e$ from $a$ to $b$ if and only if $a$'s corresponding job can be scheduled at the time slots included in $b$'s corresponding segment (let $E_2$ be these edges).
(\Rom{4}) Add a source $s$ and create a directed edge from $s$ to each vertex in $A$ (let $E_1$ be these edges).
(\Rom{5}) Add a sink $t$ and create a directed edge from each vertex in $B$ to $t$ (let $E_3$ be these edges).
(\Rom{6}) Every edge in $E_1$ has the same cost $0$ and only connects one vertex in $A$.
Let $p_j$ be $e:=(s,a)$'s capacity where $a$'s corresponding job is $j$.
(\Rom{7}) Every edge in $E_2$ has the same cost $0$, and edges that are connected with the same job vertex $j$ have the same capacity $p_j$.
(\Rom{8}) Each edge in $E_3$ only connects one vertex from $B$; let $c(\phi)$ be $e:=(b,t)$'s capacity where $\phi$ is the corresponding segment of $b$; the cost of $e$ is $1$ if there is an event occupying $\phi$ otherwise the cost of $e$ is $0$.
The goal is to find the cheapest way to send $\sum_{j\in J}p_j$ amount of flows from $s$ to $t$. 
An example is shown in \cref{fig:min-cost-flow}.

%with $s(j_1)=s(j_2)=[1,2), f(j_1)=[4,5), f(j_2)=7$, $s(e_1)=3,f(e_1)=6$, $s(e_2)=f(e_2)=7$

It is well-known that the min-cost max-flow problem can be solved in polynomial time.
The solution to min-cost max-flow has integral property as long as all capacities are integral, i.e., there exists an optimal solution to min-cost max-flow such that all flow values are integral.
\cref{lem:pseudo:agr=flow} captures the equivalence between the instance of our problem and the constructed min-cost max-flow instance.

\begin{lemma}
Given any instance $\cI$ of our problem with job set $J$ and partial solution $\cS$.
Let $\cI'$ be the constructed min-cost max-flow instance.
Then, the constructed min-cost max-flow instance has a feasible integral flow assignment $S'$ with cost $\cost(S')$ if and only if the instance $\cI$ has a job schedule $S$ with the agreement $\abs{\slot(\cS)}-\cost(S')$, where $\slot(\cS)$ is a set of time slots that are occupied by some scheduled event in $\cS$. 
\label{lem:pseudo:agr=flow}
\end{lemma}

\begin{proof}
We prove from the both directions.
\paragraph{($\Rightarrow$)}
Given a feasible integral assignment $S'$, we now construct a job schedule $S$.
For each edge $e:=(j,\phi)\in E_2$, let $f(e)$ be the total amount of flow that is assigned to edge $e$ by the solution $S'$.
The constructed job schedule is simple, i.e., for each edge $e=(j,\phi)\in E_2$, schedule $f(e)$ units of job $j$ at segment $\phi$.
We first claim that such a job schedule is a feasible schedule.
(\Rom{1}) For any job $j$, there are $p_j$ time slots that are assigned; this follows the fact that $p_j$ amount of flows are sent from vertex $j$ to $t$.
(\Rom{2}) Each time slot is assigned to at most one unit of a job; this follows the fact that the assignment $S'$ does not exceed the capacity of each edge in $E_3$.
The agreement of such a schedule is just equal to the size of $\slot(\cS)$ minus the number of time slots in $\slot(\cS)$ that are also occupied by some jobs.
This value is equal to $\cost(S')$ since all edges in $E_1,E_2$ have a cost of $0$, and edges in $E_3$ corresponding to a non-event segment also have a cost of $0$.
Hence, the agreement of the constructed schedule is equal to $\abs{\slot(\cS)}-\cost(S')$.

\paragraph{($\Leftarrow$)}
Given a feasible job schedule $S$, we now construct a feasible integral assignment $S'$ to the min-cost max-flow instance.
Note that the given job schedule $S$ consists of an assignment of jobs.
We consider each segment one by one.
For a segment $\phi$, assume that there are $q(\phi)$ units of jobs processed in this segment.
Then, we first send $q$ units of flows from $\phi$ to $t$; this is feasible since $S$ is a feasible schedule.
If a job $j$ is processed $q_j$ units in segment $\phi$, then we send $q_j$ units of flows from $j$ to $\phi$.
Finally, we send $p_j$ units of flows from $s$ to each $j$ vertex.
Since the schedule has an agreement of $\abs{\slot(\cS)}-\cost(S')$, $\cost(S')$ amounts of job units are scheduled during some segment that is included in $\slot(\cS)$.
Thus, the cost of the constructed flow has a cost of $\cost(S')$.

\end{proof}

\begin{proof}[Proof of \cref{lem:pseudo:agr}]
Given any partial solution $\cS$ and job set $J$, we construct a flow network by the method above.
By \cref{lem:pseudo:agr=flow}, we know that a feasible integral flow assignment $S$ with the minimum cost $\cost(S)$ maximizes the value of $\abs{\slot(\cS)}-\cost(S)$ since $\abs{\slot(\cS)}$ is a fixed value.
Thus, such an assignment corresponds to an optimal job schedule that produces the maximum agreement.
\end{proof}

\subsection{Submodularity of the Agreement Function}
\label{sec:pseudo:submodular}

In this section, we show that the agreement function $\agr_i(\cdot)$ is submodular for any agent $i\in A$.
To this end, we fix an agent $i\in A$ and analyze her agreement function.
For notation convenience, we drop the subscript $i$.

For convenience, each unit of a job is considered as a new job, i.e., we create $p_j$ new jobs with unit length for each job $j$.
In this way, we can assume that all jobs in $J$ have unit length.
Given any partial solution $\cS\subseteq\cL$, let $\slot(\cS)$ be a set of time slots that are occupied by scheduled events in $\cS$, so $\slot(\cS)\subseteq T$.
Given any schedule of jobs in $J$, the agreement produced by the partial solution $\cS$ and the job schedule is equal to the difference between the size of $\slot(\cS)$ and the number of time slots that are occupied by some jobs in the job schedule.
So, finding the maximum agreement produced by the partial solution $\cS$ is equivalent to finding a schedule of jobs in $J$ that uses the minimum number of time slots in $\slot(\cS)$.

This motivates us to define the following set system, which is called {\em scheduling matroid} (\cref{def:pseudo:scheduling-system}).
We shall connect the size of an independent set in the defined set system with the value of the maximum agreement later.

\begin{definition}[Scheduling Matroid]
The scheduling matroid $\cM(J):=(T,\cI)$ is defined over the time slot set $T$ for a job set $J$, where $T$ is the ground element set and $\cI$ is a collection of all independent sets.
Each job in $J$ has a release time, unit processing time, and deadline.
A set $I \subseteq T$ is an independent set if there exists a feasible schedule of $J$ only using time slots $T \setminus I$. 
% $\cI$ is a collection of all such independent sets. 
\label{def:pseudo:scheduling-system}
\end{definition}

Consider any partial solution $\cS$, let $I\subseteq\slot(\cS)$ be an independent set with the maximum size that is included in $\slot(\cS)$.
By definition, $\abs{I}$ is the maximum number of time slots in $\slot(\cS)$ such that all jobs in $J$ can be scheduled in $T\setminus I$.
Thus, the size of $I$ is equal to the maximum agreement produced by the partial solution $\cS$ since all other time slots in $\slot(\cS)\setminus I$ have to be used to ensure a feasible job schedule.
The above connection is captured by \cref{obs:pseudo:submodular}.

\begin{observation}
Given any partial solution $\cS$, 
% let $\slot(\cS)$ be the set of time slots that are occupied by some scheduled events in $\cS$.
let $I$ be a maximum independent set defined in \cref{def:pseudo:scheduling-system} that are included in $\slot(\cS)$.
Then, $\abs{I}$ is equal to the maximum agreement produced by the partial solution $\cS$.
\label{obs:pseudo:submodular}
\end{observation}

As one may observe, the maximum independent set that is included in some given element set aligns with the concept of the {\em rank} in matroid theory.
The rank function of a matroid is one of the famous submodular functions.
Hence, in the following, we aim to prove the scheduling matroid defined in \cref{def:pseudo:scheduling-system} is a matroid.
In fact, besides the submodularity, the polynomial implementation of \cref{alg:pseudo} in \cref{sec:poly} also depends on the property that the scheduling matroid is a matroid.
We start with restating the definition of a matroid and its rank function in \cref{def:pseudo:matroid}.

\begin{definition}[Matroid]
A set system $\cM:=(U,\cI)$ is a matroid if the collection $\cI$ of subsets of $U$ has the following properties: (\rom{1}) $\emptyset\in\cI$; (\rom{2}) if $A\in\cI$ and $B\subseteq A$, then $B\in\cI$; (\rom{3}) If $A,B\in\cI$ and $\abs{B}>\abs{A}$, then there exists $x\in (B\setminus A)$ such that $A\cup\set{x}\in\cI$.
Given any set $S\subseteq U$, the rank of $S$ is defined as the size of the maximum independent set included in $S$, i.e., $\rank(S):=\max_{I\in\cI:I\subseteq S}\abs{I}$.
\label{def:pseudo:matroid}
\end{definition}

\begin{lemma}
The scheduling matroid defined in \cref{def:pseudo:scheduling-system} is a matroid.    
\label{lem:pseudo:matroid}
\end{lemma}

\begin{proof}
It is not hard to see that the scheduling matroid satisfies the first and second properties stated in \cref{def:pseudo:scheduling-system}.
In the following, we focus on the third property.
We shall prove based on the bipartite graph.
Given a set of jobs $J:=\set{1,\ldots,n}$ and the time slot set $T$, we assume that the job set is feasible.
We build the following bipartite graph $G:=(A\cup B, E)$: 
(\rom{1}) for each job $j\in J$, we have one vertex $a$ in $A$;
(\rom{2}) for each time slot $t\in T$, we have vertex $b$ in $B$;
(\rom{3}) there is an edge between $a$ and $b$ if and only if $a$'s corresponding job can be scheduled at the $b$'s corresponding time slot.
See the subfigure (\rom{1}) in \cref{fig:pseudo:matroid} for an example.
Consider any time slot $T'\subseteq T$, and observe that $T'$ is an independent set if and only if there is a matching $M$ that satisfies the following two properties: (\rom{1}) $M$ matches all vertices in $A$; (\rom{2}) $M$ does not match any vertices in $B'$, where $B'$ is the corresponding vertex set of $T'$.
Given any matching $M$, let $\lt(M)\subseteq A$ and $\rt(M)\subseteq B$ be a set of vertices that are matched by $M$ in $A$ and $B$, respectively.

Now, we consider two independent sets $T_S\subseteq T$ and $T_L\subseteq T$ such that $\abs{T_S}<\abs{T_L}$.
Let $S\subseteq B$ and $L\subseteq B$ be the corresponding vertex set of $T_S$ and $T_L$, respectively.
Since $T_S$ and $T_L$ are independent sets, there exists a matching $M_S$ and $M_L$ such that $M_S$ matches all vertices in $A$ but does not match any vertices from $S$ and $M_L$ matches all vertices in $A$ but does not match any vertices from $L$. 
We distinguish three cases.
For each case, our goal is to find a vertex $v^*\in L$ and construct a new matching $M_S'$ from $M_S$ such that $M_S'$ matches all vertices in $A$ but does not match any vertices from $S\cup\set{v^*}$.
This shall imply that the defined scheduling matroid satisfies the exchange property.

Before entering the case-by-case analysis, we assume that $S\cap L=\emptyset$; otherwise, we remove all vertices in $S\cap L$ from the original bipartite graph.
This does not impact anything because both $M_S$ and $M_L$ do not use any vertices from $S\cap L$.

\paragraph{Case \Rom{1}: $\rt(M_S)\cap L=\emptyset$.}
In this case, the matching $M_S$ does not use any vertices from $L$.
Thus, we just arbitrarily pick a vertex $v^*$ from $L$ and do not need to modify $M_S$.
Because $M_S$ is the desired matching, i.e., $M_S$ matches all vertices from $A$ but does not match any vertices from $S\cup\set{v^*}$, where $v^*\in L$.

\paragraph{Case \Rom{2}: $\rt(M_L) \subseteq S$.}
In this case, all vertices in $\rt(M_L)$ are in $S$.
This implies that the size of $S$ is at least $\abs{\rt(M_L)}$, i.e., $\abs{S}\geq\abs{\rt(M_L)}=\abs{A}$.
Since $\abs{L}>\abs{S}$, we have $\abs{L}\geq \abs{A}+1$.
This implies that there exists at least one vertex in $L$ that is not used by $M_S$ because $M_S$ matches at most $\abs{A}$ vertices from $B$.
Suppose that such a vertex is $v^*$.
Hence, we know that $v^*$ is not used by $M_S$ and $v^*\in L$.
Note that $v^*\notin S$ since $S\cap L=\emptyset$.
Thus, $v^*$ and $M_S$ are the desired vertex and matching.
Because $M_S$ matches all vertices from $A$ and does not use any vertices from $S\cup\set{v^*}$, where $v^*\in L$.

\paragraph{Case \Rom{3}: $\rt(M_S)\cap L\ne\emptyset$ and $\rt(M_L)\setminus S\ne\emptyset$.}
In this case, we know the following facts: (\rom{1}) $M_S$ uses some vertices from $L$; (\rom{2}) $M_L$ uses some vertices that are not in $S$.
In the case where there exists some vertex $v^*$ in $L$ that is not used by $M_S$, $v^*$ and $M_S$ are the desired vertex and matching; the reason is the same as the second case.
Thus, it is safe to assume that $M_S$ matches all vertices in $L$, i.e., $L\subseteq \rt(M_S)$.
Observe that we have the following two facts: (\rom{1}) the size of $\rt(M_L)\setminus S$ is at least $\abs{A}-\abs{S}$; (\rom{2}) the size of $\rt(M_S)$ is at least $\abs{L}$.
This implies that $\abs{L}+\abs{\rt(M_L)\setminus S}>\abs{A}$ since $\abs{L}>\abs{S}$.
Thus, there exists at least one vertex in $A$ (denoted by $u^*$) such that the following two statements are true: (\rom{1}) $u^*$ is matched by some vertex $v^*_s$ in $L$ in $M_S$; (\rom{2}) $u^*$ is matched by some vertex $v^*_l$ in $\rt(M_L)\setminus S$ in $M_L$.
Now, we construct a new matching $M_S'$ from $M_S$: replacing the edge $(u^*,v^*_s)$ with the edge $(u^*,v^*_l)$, i.e., $M_S'=M_S\setminus \set{(u^*,v^*_s)} \cup\set{(u^*,v^*_l}$.
We obtain the desired vertex $v^*_s$ and the desired matching $M_S'$.
Because $M_S'$ matches all vertices from $A$ but does not use any vertices from $S\cup\set{v^*_s}$, where $v^*_s\in L$.
See \cref{fig:pseudo:matroid} for an example.

\begin{figure}[htb]
    \centering
    \includegraphics[width=15cm]{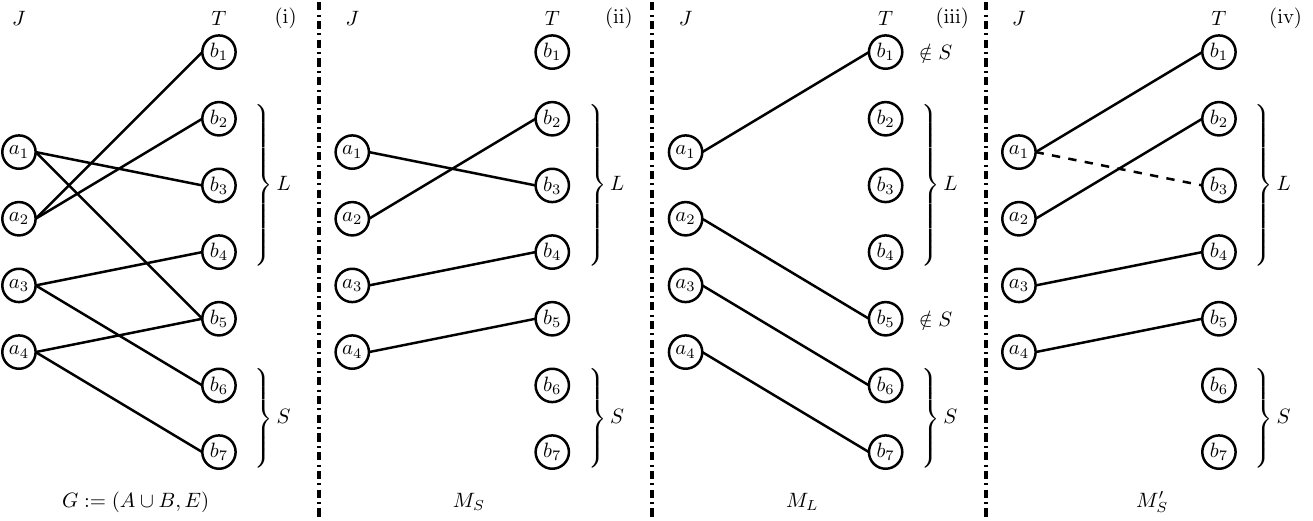}
    \caption{Illustration of Case (\Rom{3}). The subfigure (\rom{1}) shows the original bipartite graph $G:=(A\cup B, E)$, where $A$ corresponds to the job set and $B$ corresponds to the time slot set. The large vertex set $L=\set{b_2,b_3,b_4}$ and the small vertex set $S=\set{b_6,b_7}$. The subfigure (\rom{2}) shows a matching $M_S$ that does not use any vertices from $S$. The subfigure (\rom{3}) shows a matching $M_L$ that does not use any vertices from $L$. We can prove that $\abs{L}+\abs{\rt(M_L)\setminus S}>\abs{A}$. In the example, $\rt(M_L)\setminus S=\set{b_1,b_5}$ which is shown in the subfigure (\rom{3}). This inequality implies that a vertex exists in $A$ that forms an alternative path. In the example, such a vertex is $a_1$, where $a_1$ is matched with $b_3\in L$ in $M_S$ and $a_1$ is matched with $b_1\in \rt(M_L)\setminus S$. Thus, the path $(b_1,a_1,b_3)$ forms an alternative path shown in the subfigure (\rom{4}). By rematching along the alternative path, we match $a_1$ with other vertices that are not in $S$ and thus release a vertex from $L$.}
    \label{fig:pseudo:matroid}
\end{figure}   
\end{proof}

\begin{proof}[Proof of \cref{lem:pseudo:submodular}]
By \cref{lem:pseudo:matroid}, we know that the defined scheduling matroid is a matroid.
By \cref{obs:pseudo:submodular}, we know that, for any partial solution $\cS$, the value of $\agr(\cS)$ is equal to the rank of $\st(\cS)$, i.e., $\agr(\cS)=\rank(\st(\cS))$.
Note that $\st(\cdot)$ can be considered as a coverage function, i.e., it maps any subset of $\cL$ to a subset of $T$.
It is well known that the ranking function of a matroid is submodular, e.g., see~\cite[Lemma 5.1.3]{lau2011iterative} for a formal proof.
To prove the submodularity of $\agr(\cdot)$, we show the following inequality, which is the definition of a submodular function: for any $\cS,\cH\subseteq \cL$,
$$
\agr(\cS) + \agr(\cH) \geq \agr(\cS\cup\cH) + \agr(\cS\cap \cH).
$$
The above inequality holds mainly because of \cref{lem:pseudo:matroid} and \cref{obs:pseudo:submodular}.
We have the following inequalities:
\begin{align*}
\agr(\cS) + \agr(\cH) &= \rank(\st(\cS)) + \rank(\st(\cH)) \tag{\cref{obs:pseudo:submodular}} \\
&\geq \rank(\st(\cS)\cup\st(\cH)) + \rank(\st(\cS)\cap \st(\cH)) \tag{Submodularity of $\rank$} \\
&=\rank(\st(\cS\cup\cH)) + \rank(\st(\cS)\cap \st(\cH)) \tag{$\st(\cS\cup\cH)=\st(\cS)\cup\st(\cH)$}\\
&\geq \rank(\st(\cS\cup\cH)) + \rank(\st(\cS\cap\cH)) \tag{$\st(\cS\cap\cH)\subseteq \st(\cS)\cap\st(\cH)$} \\
&=\agr(\cS\cup\cH) + \agr(\cS\cap\cH) \tag{\cref{obs:pseudo:submodular}} 
\end{align*}
Thus, the agreement function $\agr:2^{\cL}\to\N_{\geq 0}$ is a submodular function.
\end{proof}

\subsection{Submodular Maximization with Group Constraints and Imperfect Oracle}
\label{sec:pseudo:group}

In this section, we show greedily picking an element with a ``good'' marginal value from the unselected groups is a $\frac{1}{\alpha+1}$-approximate algorithm for SMGC under an $\alpha$-approximate oracle $g(\cdot,\cdot)$ (see \cref{alg:pseudo:group}).
% We cannot find the element with a maximum marginal value because of $\alpha$-approximate oracle.
We cannot find the element with the exact maximum marginal value because the oracle is $\alpha$-approximate.
% The formal description of the algorithm can be found in \cref{alg:pseudo:group} in \cref{app:sec:pseudo:group}.
When $\alpha=1$, \cref{alg:pseudo:group} is $\frac{1}{2}$-approximate, which aligns with the result for submodular maximization subject to matroid constraint stated in ~\cite{horel2015notes}.
But our proof only uses the monotone and submodular properties and thus simplifies the proof in~\cite{horel2015notes}.
For $\alpha>1$, our result generalizes the approximation ratio stated in~\cite{DBLP:conf/approx/ChekuriK04} where they also consider the $\alpha$-approximate oracle but only focus on coverage functions.

\begin{algorithm}[H]
\caption{Natural Greedy for SMGC}
\label{alg:pseudo:group}
\begin{algorithmic}[1]
\Require The ground element set $U$ and its group collection $G_1,\ldots,G_{\kappa}$; A monotone submodular function $f:2^U\to \R_{\geq 0}$; The capacity $k\in\N_{\geq 0}$.
\Ensure A subset $S\subseteq U$ with $\abs{S}\leq k$.
\State $S\leftarrow\emptyset$.
\While{$\abs{S}\leq k$}
\State $H\leftarrow\emptyset$.
\For{each unselected group $G_i$}
\State $a_i\leftarrow g(S,G_i)$.
\State $H\leftarrow H \cup\set{a_i}$.
\EndFor
\State $a^*\leftarrow\max_{b\in H}f(S\cup\set{b})-f(S)$.
\State $S\leftarrow S\cup\set{a^*}$.
\EndWhile
\State \Return $S$.
\end{algorithmic}
\end{algorithm}

\begin{proof}[Proof of \cref{lem:pseudo:group}]
Let $a_1,a_2,\ldots,a_k$ be the elements selected by \cref{alg:pseudo:group}.
We assume that these elements are added to the solution $S$ following this order, and $a_i$ is the element from the $i$-th group.
This can be achieved by reordering the index of the groups.
Define $S^{j}:=\set{a_1,\ldots,a_j}$ as the first $j$ elements selected by \cref{alg:pseudo:group}.
Note that $S^{0}=\emptyset$ and $S^{k}=S$.
Let $O^*:=\set{o^*_1,\ldots, o^*_k}$ be the optimal solution, where $o^*_i$ comes from the $i$-th group.
Given any subset $H\subseteq U$, we use $f_H(\cdot)$ to denote the marginal function, i.e., $f_H(T):=f(H\cup T)-f(H)$.

The analysis mainly relies on the following simple claim (\cref{clm:pseudo:group}), which holds because of the submodularity.

\begin{claim}
For any $j\in\set{1,2,\ldots,k}$, we have $f(S^{j-1}\cup\set{a_j})-f(S^{j-1})\geq\frac{1}{\alpha} f_{S}(o^*_j)$.
\label{clm:pseudo:group}
\end{claim}
\begin{proof}
We start with a property of the marginal function of a submodular function.
\begin{property}
For any $L,S,T\subseteq U$ with $S\subseteq L$, we have $f_S(T)\geq f_L(T)$.   
\label{pro:submodular}
\end{property}

Since \cref{alg:pseudo:group} chooses $a_j$ at the $j$-th iteration, we know that $g(S^{j-1},G_j)=a_j$.
Thus, by the definition of the oracle $g$, we have:
$$
f(S^{j-1}\cup\set{a_j})-f(S^{j-1})\geq \frac{1}{\alpha}\cdot \max_{e\in G_j}f(S^{j-1}\cup\set{e})-f(S^{j-1}).
$$
Since $o^*_j$ is also in $G_j$, we have: 
$$
\max_{e\in G_j}f(S^{j-1}\cup\set{e})-f(S^{j-1})\geq f(S^{j-1}\cup\set{o^*_j})-f(S^{j-1}).
$$
Combining the above two inequalities, we have:
$$
f(S^{j-1}\cup\set{a_j})-f(S^{j-1})\geq \frac{1}{\alpha}\cdot \left( f(S^{j-1}\cup\set{o^*_j})-f(S^{j-1}) \right).
$$
By \cref{pro:submodular}, we have $f(S^{j-1}\cup\set{o^*_j})-f(S^{j-1}) \geq f_S(o^*_j)$ since $S^{j-1}\subseteq S$ for all $j\in[k]$.
Thus, we have:
$$
f(S^{j-1}\cup\set{a_j})-f(S^{j-1})\geq \frac{1}{\alpha}\cdot \left( f(S^{j-1}\cup\set{o^*_j})-f(S^{j-1}) \right) \geq \frac{1}{\alpha}\cdot\left( f(S\cup\set{o^*_j})-f(S)\right) = \frac{1}{\alpha}\cdot f_S(o^*_j).
$$
Thus, \cref{clm:pseudo:group} holds.
\end{proof}

Based on \cref{clm:pseudo:group}, we have the following inequalities:
\begin{align*}
f(S)=\sum_{j=1}^{k} \left(f(S^{j-1}\cup\set{a_j})-f(S^{j-1}) \right) 
&\geq  \frac{1}{\alpha}\cdot \left (\sum_{j=1}^{k}f_S(o^*_j)\right) \tag{\cref{clm:pseudo:group}} \\ 
&\stackrel{(\rom{1})}{\geq} \frac{1}{\alpha}\cdot f_S(O^*) \\
&= \frac{1}{\alpha}\cdot\left( f(S\cup O^*)-f(S) \right) \\
&\stackrel{(\rom{2})}{\geq} \frac{1}{\alpha}\cdot\left( f(O^*)-f(S) \right). 
\end{align*}
The inequality (\rom{1}) is due to the fact that the marginal function of a submodular function is subadditive\footnote{A function $f$ is subadditive if $f(L\cup T)\leq f(L)+f(T)$ for any $L,T\subseteq U$.}.
The inequality (\rom{2}) is due to the fact that $f$ is a monotone function.
Thus, we have $f(S)\geq\frac{1}{\alpha+1}f(O^*)$ which proves \cref{lem:pseudo:group}.
\end{proof}

\section{An Algorithmic Framework for Arbitrary T}\label{sec:poly}

In this section, we consider the case where 
$\abs{T}$ is an arbitrary value and present an algorithmic framework for this case.
Given any approximation algorithm for the one-event instance, the proposed algorithmic framework rooted in \cref{alg:pseudo} is able to extend the given approximation algorithm to the general case by only losing a small constant factor on the approximation ratio.
The formal description can be found in \cref{thm:general-T}.
The one-event instance is a special case of \PSBA/ where there is only one event that needs to be scheduled; we use \PSBA/ with $\abs{E}=1$ to denote this case.

\begin{theorem}
Given any $\frac{1}{\alpha}$-approximate algorithm ($\alpha\geq 1$) running in polynomial time for \PSBA/ with $\abs{E}=1$, there is a polynomial time algorithm that achieves $\frac{1}{\alpha+1}$-approximation for general instances of \PSBA/. 
\label{thm:general-T}
\end{theorem}

Before entering the algorithmic idea for \cref{thm:general-T}, we first discuss the issue that converts \cref{alg:pseudo} to a polynomial time algorithm when $T$ is arbitrary.
As is typical, one might expect to use the standard technique to cut the whole time span $T$ into a polynomial number of segments. 
Then, one might be able to prove that the algorithm does not lose too much on this discrete-time span (usually a $\epsilon$ factor in many classical scheduling problems such as~\cite{DBLP:conf/soda/AntoniadisGKK20,DBLP:journals/siamcomp/Li20}).
Unfortunately, this is not the case for our problem since a slight movement of an event schedule may change agents' agreement from positive to zero and thus lead to an unbounded approximation ratio. 

\paragraph{Algorithmic Ideas.}
To design an algorithmic framework that achieves \cref{thm:general-T}, our idea is to release the full power of \cref{alg:pseudo} and \cref{lem:pseudo:group}.
Recall that in \cref{sec:poly}, we consider the polynomially bounded $\abs{T}$, which is equivalent to assuming that we have a $1$-approximate oracle for SMGC in~\cref{def:pseudo:submodular-maximization}.
By \cref{lem:pseudo:group}, as long as we have an $\alpha$-approximate oracle, \cref{alg:pseudo} with line~\ref{line:pseudo:max} replaced (\cref{alg:poly}) is a $\frac{1}{\alpha+1}$-approximate algorithm.
Thus, in this section, we aim to design such an $\alpha$-approximate oracle to replace line~\ref{line:pseudo:max} of \cref{alg:pseudo}.
To this end, we need to build an algorithm $\mathrm{ALG}$ (\cref{alg:poly:position}) such that $\mathrm{ALG}$ satisfies the following two properties: (\rom{1}) $\mathrm{ALG}$ runs in polynomial time (\cref{obs:poly:position:running-time}); (\rom{2}) given any partial solution $\cS$ and an event $e$, $\mathrm{ALG}$ returns a position for the event $e$ such that the resulting schedule is $\frac{1}{\alpha}$-approximate (\cref{lem:poly:position}).
% Thus, in this section, we aim to design such an $\alpha$-approximate oracle to replace line~\ref{line:pseudo:max} of \cref{alg:pseudo}.
% To this end, we need to build an algorithm $\mathrm{ALG}$ (\cref{alg:poly:position}) such that $\mathrm{ALG}$ satisfies the following two properties (\cref{lem:poly:position}): (\rom{1}) $\mathrm{ALG}$ runs in polynomial time; (\rom{2}) given any partial solution $\cS$ and an event $e$, $\mathrm{ALG}$ returns a position for the event $e$ such that such a schedule is $\frac{1}{\alpha}$-approximate.

We shall build the polynomial time oracle based on the given $\frac{1}{\alpha}$-approximate algorithm for \PSBA/ with $\abs{E}=1$.
% Given a $\frac{1}{\alpha}$-approximate algorithm for \PSBA/ with $\abs{E}=1$, we design a desired polynomial-time oracle based on such an approximation algorithm.
To achieve this, the main obstacle is that the given one-event instance algorithm only works for the case where there is no partial solution; this disagrees with the requirement of the oracle.
Our idea to fix this issue is to construct an equivalent pure one-event instance so that when we run $\frac{1}{\alpha}$-approximate algorithm on the constructed one-event instance, it is able to return a good position for the instance with the partial solution.
This method works mainly due to the exchange property of the scheduling matroid.

Formally, we use $\onet(\cdot,\cdot)$ to represent the given $\frac{1}{\alpha}$-approximate algorithm for \PSBA/ with $\abs{E}=1$.
The algorithm $\onet(A,e^*)$ takes the agent set $A$ (including their job sets) and an event $e^*$ as the input; its output is a time slot $t^*$ that indicates the scheduled position for the given event.
By our assumption, the returned time slot $t^*$ satisfies the following property: 
$
\sum_{i\in A}\agr_i(\set{(e^*,t^*)}) \geq \frac{1}{\alpha}\cdot \max_{t\in T}\sum_{i\in A}\agr_i(\set{(e^*,t)}).
$
% Note that, in this section, we aim to design an algorithm (\cref{alg:poly:position}), denoted by $\gp(\cdot,\cdot)$, to replace line~\ref{line:pseudo:max} of \cref{alg:pseudo}.
Let $\gp(\cdot,\cdot)$ be the desired approximate oracle, and $\gp(\cS,e^*)$ shall use the given $\frac{1}{\alpha}$-approximate one-event algorithm as a subroutine. It takes two parameters as the input: the given partial solution $\cS$ and an event $e^*$, and will return a time slot that indicates the scheduled time of the given event.
Recall that, given a partial solution $\cS$, we need to construct an equivalent pure one-event instance $A'$ such that the time slot $t^*$ returned by $\onet(A',e^*)$ is a good position for $e^*$, i.e., 
$
\sum_{i\in A}\agr_i(\cS\cup\set{(e^*,t^*)}) \geq \frac{1}{\alpha} \cdot \max_{t\in T}\sum_{i\in  A}\agr_i(\cS\cup\set{(e^*,t)}).
$
The above inequality shall imply that $\gp(\cdot,\cdot)$ is exactly the same as the $\alpha$-approximate oracle $g$ in \cref{def:pseudo:submodular-maximization}, which leads to a $\frac{1}{\alpha+1}$-approximate algorithm by \cref{lem:pseudo:group}.

In the following, we focus on how to construct such an equivalent instance $A'$.
To this end, we start with an important property of the scheduling matroid defined in \cref{def:pseudo:scheduling-system}; see \cref{lem:poly:position:matroid-property}.
Recall that the scheduling matroid is a matroid, and it is defined over the whole time slot set and a job set.

\begin{lemma}
For a scheduling matroid $\cM(J):=(T,\cI)$ defined in \cref{def:pseudo:scheduling-system}, consider a time slot set $S\subseteq T$ and let $F\in\cI$ be a maximum independent set that is included in $S$.
For each time slot in $F$, create a rigid job and let $J'$ be all these rigid jobs.
The job set $J\cup J'$ defines a new scheduling matroid, denoted by $\cM(J\cup J'):=(T,\cI')$.
Consider a time slot set $H\subseteq T$ and let $B\in\cI'$ be a maximum independent set that is included in $H$.
Then, $F\cup B\in\cI$ is a maximum independent set in $\cM(J)$ that is included in $S\cup H$.
\label{lem:poly:position:matroid-property}
\end{lemma}

\begin{proof}
To prove \cref{lem:poly:position:matroid-property}, it is sufficient to show the following three claims: (\rom{1}) $F\cup B\subseteq S\cup H$; (\rom{2}) $F\cup B\in \cI$ is an independent set of $\cM(J)$; (\rom{3}) $F\cup B$ is a maximum independent set.
The first claim is clearly true since $F\subseteq S$ and $B\subseteq H$.
To show claim (\rom{2}), we prove that there exists a feasible schedule for $J$ only using time slots in $T\setminus (F\cup B)$.
This is true because $B\in\cI'$ is an independent set; so, there exists a feasible schedule $\Gamma$ for $J\cup J'$ only using time slots in $T\setminus B$.
Note that in $\Gamma$, no job from $J$ uses a time slot from $F$ since every time slot of $F$ is occupied by a job in $J'$.
And also, no job from $J$ uses a time slot from $B$.
Thus, after removing all jobs from $J'$, $\Gamma$ becomes a feasible schedule for $J$ and $\Gamma$ does not use any time slot from $F\cup B$.
Hence, $F\cup B\in\cI$ is an independent set of matroid $M(J)$.
For claim (\rom{3}), we prove by contradiction and thus assume that there exists another independent set $K\in\cI,K\subseteq S\cup H$ with $\abs{K}>\abs{F\cup B}$.
By the exchange property, we know that there exists an element $x\in K\setminus (F\cup B)$ such that $\set{x}\cup F\cup B$ is an independent set.
Note that $x$ is either in $S$ or $H$ since $K\subseteq S\cup H$.
If $x\in S$, then $F\cup\set{x}$ is an independent set of $S$; so, $F$ is not a maximum independent set of $S$, a contradiction.
If $x\in H$, then $B\cup\set{x}$ is an independent set of $H$; so, $B$ is not a maximum independent set of $H$, a contradiction.
Hence, $F\cup B$ is a maximum independent set in $\cM(J)$ that is included in $S\cup H$.
\end{proof}

\begin{figure}[htb]
    \centering
    \includegraphics[width=10cm]{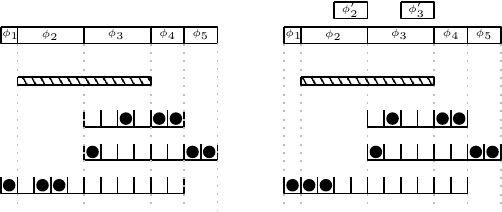}
    \caption{Illustration of time slots aggregation (line~\ref{line:aggregate} of \cref{alg:poly:position}). The subfigure (i) presents an optimal job schedule computed by the min-cost max-flow algorithm. If we do not gather the agreement time slots, it is possible that the agreement time slots are separated by the job's processing. In subfigure (ii), such time slots producing agreements are aggregated to form agreement segments. Each segment can have at most one agreement segment.}
    \label{fig:aggrslot}
\end{figure}

Given any partial solution $\cS$, recall that a maximum independent set included in $\slot(\cS)$ is a set of time slots with the maximum size that produces the agreement (\cref{obs:pseudo:submodular}). We call these time slots {\em agreement time slots}.
Thus, \cref{lem:poly:position:matroid-property} actually suggests a ``stability'' property of these agreement time slots.
In other words, consider two partial solutions $\cS$ and $\cH$.
The agreement time slot of the partial solution $\cS$ is a subset of the agreement time slot of the partial solution $\cS\cup\cH$.
This crucial stable property motivates the pure one-event instance construction method, where we shall replace all agreement time slots of partial solution $\cS$ with many rigid jobs.
In this way, we eliminate the impact of the partial solution $\cS$ and focus on finding the maximum agreement of the partial solution $\cH$.
Then, by \cref{lem:poly:position:matroid-property}, the union of $\cS$'s and $\cH$'s agreement time slots shall be the maximum agreement time slots of the whole solution $\cS\cup\cH$.
The only issue is that we need to be careful when creating rigid jobs since the number of constructed rigid jobs is required to be polynomial.
This requirement can be satisfied by aggregating all agreement time slots in each segment when computing the min-cost flow (\cref{alg:poly:position}).
An example is shown in \cref{fig:aggrslot}.

\begin{algorithm}[htb]
\caption{\texttt{GoodPosn}$(\cdot,\cdot)$}
\label{alg:poly:position}
\begin{algorithmic}[1]
\Require A partial solution $\cS$; An unscheduled event $e^*$. 
\Ensure A good position $t^*$ for event $e^*$.
\For{each agent $i\in A$}
\State Compute a schedule for $J_i$ by min-cost max-flow.
\For{each segment in $\set{\phi\in\Phi\mid \phi\subseteq \slot(\cS)}$}
\State Aggregate all agreement time slots forming an agreement segment $\phi'$.
\label{line:aggregate}
\State Create a rigid job $j'$ for $\phi'$.
\EndFor
\State $J_i\leftarrow J_i \cup \set{\text{all created rigid jobs}}$.
\EndFor
\State \Return $t^*\leftarrow\texttt{OneEvent}(A,e^*)$.
\end{algorithmic}
\end{algorithm}

\begin{observation}
\cref{alg:poly:position} runs in polynomial time, i.e., \cref{alg:poly:position} creates a polynomial number of rigid jobs.
\label{obs:poly:position:running-time}
\end{observation}
\begin{proof}
For each segment, we have at most one agreement segment. 
Thus, the total number of created agreement segments is at most $\abs{\Phi}$.
\end{proof}
    
\cref{lem:poly:position} states that \cref{alg:poly:position} $\gp(\cdot,\cdot)$ is a valid $\alpha$-approximate-oracle for the submodular maximization problem stated in \cref{def:pseudo:submodular-maximization}.

\begin{lemma}
Given any partial solution $\cS$ and an unscheduled event $e^*$, let $t^*$ be the time slot returned by \cref{alg:poly:position}.   
Then, we have 
$
\sum_{i\in A}\agr_i(\cS\cup\set{(e^*,t^*)}) \geq \frac{1}{\alpha} \cdot \max_{t\in T}\sum_{i\in  A}\agr_i(\cS\cup\set{(e^*,t)}).
$
\label{lem:poly:position}
\end{lemma}

\begin{proof}
% To prove \cref{lem:poly:position}, it is sufficient to prove that for all $i\in A$, we have:
% $$
% \agr_i(\cS\cup\set{(e^*,t^*)}) \geq \frac{1}{\alpha} \cdot \max_{t\in T}\agr_i(\cS\cup\set{(e^*,t)}).
% $$
Recall that $t^*$ is returned by $\onet(A,e^*)$, where each agent $i\in A$ has an additional rigid job set $J_i'$.
In order to distinguish, we assume that the job set of agent $i'$ is $J_i\cup J_i'$, and agent $i$'s job set is still $J_i$.
Let $A'$ be the set of agent $i'$, i.e., $A':=\set{i'}_{i\in A}$.
Each job in $J_i'$ indicates the time slots that produce agreement based on the partial solution $\cS$.
Let $\slot(J_i')$ be a set of time slots that are included in some jobs in $J_i'$.
Note that $\abs{\slot(J_i')}$ is equal to the maximum agreement of agent $i$ when the given partial solution is $\cS$.
By the definition of $\onet(\cdot,\cdot)$, we have:
\begin{equation}
\sum_{i'\in A'}\agr_{i'}(\set{(e^*,t^*)})\geq \frac{1}{\alpha}\cdot \max_{t\in T}\sum_{i'\in A'}\agr_{i'}(\set{(e^*,t)}).    
\label{equ:poly:position:key-1}
\end{equation}
Now, we focus on an agent $i\in A$, and define a scheduling matroid $\cM(J_i):=(T,\cI)$.
Note that $\slot(J_i')\in\cI$ is a maximum independent set of $\slot(\cS)$.
Based on the time slot set $\slot(J_i')$, we define a scheduling matroid $\cM(J_i\cup J_i'):=(T,\cI')$.
Consider any $t\in T$, we use $\cH$ to denote the partial solution $\set{(e^*,t)}$ for notation convenience.
Let $B\in\cI'$ be a maximum independent set of $\slot(\cH)$; so, we have $\agr_{i'}(\cH)=\abs{B}$.
By \cref{lem:poly:position:matroid-property}, we know that $B\cup \slot(J_i')\in\cI$ is a maximum independent set of $\slot(\cS)\cup\slot(\cH)$; so, $\agr_i(\cS\cup\cH)=\abs{B}+\abs{\slot(J_i')}$.
Hence, for any $t\in T$ and $i\in A$, we have the following key equality:
\begin{equation}
\agr_i(\cS\cup\set{(e^*,t)}) = \abs{\slot(J_i')}+\agr_{i'}(\set{(e^*,t)}).
\label{equ:poly:position:key-2}
\end{equation}

After adding $\sum_{i'\in A'}\abs{\slot(J_i')}$ to both sides of \cref{equ:poly:position:key-1}, we have:
\begin{align*}
\sum_{i\in A}\agr_{i}(\cS\cup\set{(e^*,t^*)})
&= \sum_{i'\in A'}\abs{\slot(J_i')}+\sum_{i'\in A'}\agr_{i'}(\set{(e^*,t^*)}) \tag{\cref{equ:poly:position:key-2}} \\
&\geq \sum_{i'\in A'}\abs{\slot(J_i')}+ \frac{1}{\alpha}\cdot \max_{t\in T}\sum_{i'\in A'}\agr_{i'}(\set{(e^*,t)}) \tag{\cref{equ:poly:position:key-1}} \\
&\geq \frac{1}{\alpha} \cdot\left( \sum_{i'\in A'}\abs{\slot(J_i')}+ \max_{t\in T}\sum_{i'\in A'}\agr_{i'}(\set{(e^*,t)})\right)\tag{$\alpha\geq 1$} \\
&\geq \frac{1}{\alpha} \cdot \max_{t\in T}\sum_{i'\in A'}\left( \abs{\slot(J_i')}+\agr_{i'}(\set{(e^*,t)}) \right) \\
&=\frac{1}{\alpha}\cdot\max_{t\in T}\sum_{i\in A}\agr_i(\cS\cup\set{(e^*,t)}) \tag{\cref{equ:poly:position:key-2}}
\end{align*}
Therefore, \cref{lem:poly:position} holds.
\end{proof}

Now, we are ready to provide the complete algorithm for arbitrary $T$.
\cref{alg:poly} just replaces line~\ref{line:pseudo:max} of \cref{alg:pseudo} with \cref{alg:poly:position}.

\begin{algorithm}[H]
\caption{Algorithmic Framework for Arbitrary $T$.}
\label{alg:poly}
\begin{algorithmic}[1]
\Require An instance of \PSBA/;
\Ensure A schedule $\cS$ of all events in $E$.
\State $\cS\leftarrow\emptyset$.
\While{there exists an unscheduled event}
\State $\cH\leftarrow\emptyset$.
\For{each unscheduled event $e$}
\State $t\leftarrow\gp(\cS,e)$.
\State $\cH\leftarrow \cH\cup \set{(e,t)}$.
\EndFor
\State $(e^*,t^*)\leftarrow \argmax_{(e,t)\in\cH} F(\cS\cup\set{(e,t)})-F(\cS)$.
\State $\cS\leftarrow \cS\cup\set{(e^*,t^*)}$.
\Comment{Schedule event $e$ at time slot $t$.}
\EndWhile
\State \Return $\cS$.
\end{algorithmic}
\end{algorithm}

\begin{proof}[Proof of \cref{thm:general-T}]
\cref{alg:poly:position} is based on the $\frac{1}{\alpha}$-approximate algorithm for \PSBA/ with $\abs{E}=1$.
Thus, if there exists such an approximation algorithm running in polynomial time, then \cref{alg:poly:position} exists, and it runs in polynomial time by \cref{obs:poly:position:running-time}.
By \cref{lem:poly:position}, given any partial solution $\cS$ and an event $e$, \cref{alg:poly:position} is able to find a position $t$ for event $e$ such that schedules $e$ at $t$ is a $\frac{1}{\alpha}$-approximate solution.
Thus, \cref{alg:poly:position} matches the definition of an $\alpha$-approximate oracle.
By \cref{lem:pseudo:group}, we know that \cref{alg:poly} is a $\frac{1}{\alpha+1}$-approximate algorithm.
\end{proof}

\section{An Optimal Algorithm for One-event Instance}

\label{sec:poly:one-event}
In this section, we focus on the case where there is only one event $e$ to be scheduled, and we try to find a position for $e$ that produces the maximum agreement. 
Formally, we prove the following result (\cref{thm:one-event}).

\begin{theorem}
Given any instance of \PSBA/ with $\abs{E}=1$, there is an algorithm running in polynomial time that returns an optimal position for the event.
\label{thm:one-event}
\end{theorem}

Our idea is to break down such a one-event scheduling problem into a polynomial number of subproblems, each of which is a special instance $\hcI$ with only one agent who has one multi-interval job and a bunch of rigid jobs, and these jobs' intervals are non-overlap.
We call such a special instance $\hcI$ a {\em core instance}.
We first design a polynomial algorithm for the core instance in \cref{sec:ins}. 
Then we show in \cref{sec:gen} how to compute the optimal event schedule for the general one-event instance by running the algorithm in \cref{sec:ins}.

\subsection{Core Instance}
\label{sec:ins}

We first formally define the core instance.
The core instance $\hcI$ consists of one event $e$ and one agent $i$. 
In this section, we focus on agent $i$, and we shall drop the subscript $i$ for notation convenience.
The agent $i$'s job set $J=\set{j^*}\cup J'$ contains one multi-interval job $j^*$ and a set of rigid jobs $J'$. 
Let $p^*$ be the processing time of job $j^*$.
Let $\set{I^*_1,\ldots,I^*_{\kappa}}$ and $\set{I_{j'}}$ be the job intervals of $j^*$ and a rigid job $j'\in J'$, respectively.
Each job interval $I$ is considered as a set of time slots, i.e., $I\subseteq T$.
The job intervals of $j^*$ and each job interval of a rigid job in $J'$ have no overlap with each other. 
To find an optimal scheduled position for event $e$, we aim to draw the figure of the agreement function $\agr(\set{(e,t)})$, i.e., the maximum agreement produced by the solution that schedules event $e$ at $t$.
In other words, the x-axis is the time, and the y-axis is the value of the agreement function.
We use $g:T\to\N_{\geq 0}$ to denote the value of maximum agreement at each time slot, i.e., $g(t):=\agr(\set{(e,t)})$.
It is not hard to see that $g(t)$ is piecewise linear, and thus, to know the value of $g(t)$ at each $t\in[T]$, we only need to compute its turning points within polynomial time.

Let $S\subseteq T$ be a set of time slots.
We define a new function $h(\cdot,\cdot):(2^T,T)\to\N_{\geq 0}$ that indicates the number of time slots that are covered by both the event and $S$, i.e., $h(S,t):=\abs{S\cap \slot(\set{e,t})}$.
For any $t\in T$, the value of $h(S,t)$ and $g(t)$ has a strong connection; this can be captured by the following lemma (\cref{lem:poly:core-instance:g-h}).
The connection is mainly due to the fact that all job intervals do not intersect.

\begin{lemma}
For any $t\in T$, $g(t)=l(e)-\sum_{j'\in J'}h(I_{j'},t)-\max\{p^*-(|\bigcup_{i\in[\kappa]}I^*_i|-\sum_{i\in[\kappa]}h(I_{i}^*,t)),0\}$.    
\label{lem:poly:core-instance:g-h}
\end{lemma}

\begin{proof}
The agreement $g(t)$ is equal to the length of the event ($l(e)$) minus busy time slots that are included in $\slot(\set{(e,t)})$, where busy time slots are those who are occupied due to some jobs' processing.
Every time slot in $\bigcup_{j'\in J'}I_{j'}$ is busy since all jobs in $J'$ are rigid jobs.
Thus, we need to subtract those busy time slots that are also included in $\slot(\set{(e,t)})$; this part is equal to $\sum_{j'\in J'}h(I_{j'},t)$.
For those time slots that are included in both $\slot(\set{(e,t)})$ and $\bigcup_{i\in[\kappa]}I^*_i$, they may not be busy since $j^*$ is a flexible job.
To maximize the agreement, the optimal solution must use the minimum number of time slots that are occupied by event $e$; this value is equal to $\max\{p^*-(|\bigcup_{i\in[\kappa]}I^*_i|-\sum_{i\in[\kappa]}h(I_{i}^*,t)),0\}$ where $|\bigcup_{i\in[\kappa]}I^*_i|-\sum_{i\in[\kappa]}h(I_{i}^*,t))$ is the number of time slots occupied by $j^*$ but not by the event.
Thus, \cref{lem:poly:core-instance:g-h} holds.
\end{proof}

Based on \cref{lem:poly:core-instance:g-h}, we know that the function $g(\cdot)$ is able to be determined by a polynomial number of $h(\cdot,\cdot)$ functions.
Thus, the picture of the function $g(\cdot)$ can be produced by the picture of the function $h(\cdot,\cdot)$ in a polynomial number of steps via some simple mappings; see \cref{lem:poly:core-instance:agrfig} for a formal description.
Therefore, to draw $g(\cdot)$, it is sufficient to show that all turning points of the function $h(\cdot,\cdot)$ can be computed in polynomial time.
To this end, we consider an arbitrary time slot set $S$, and prove that $h(S,\cdot):T\to\N_{\geq 0}$ can be drawn in polynomial time; see \cref{lem:poly:core-instance:h}.
For notation convenience, we use $h_S(\cdot)$ to stand for $h(S,\cdot)$.

\begin{lemma}\label{lem:poly:core-instance:h}
    Given an interval $S\subseteq T$, the function $h_S:T\to\N_{\geq 0}$ can be drawn in polynomial time.
\end{lemma}

\begin{proof}
Recall that for any $t\in T$, $h_S(t)$ is equal to the number of time slots that are occupied by both $S$ and the event.
Observe that the function $h_S$ is also piecewise linear, and thus, we only need to find its turning points. 
Define $T':=\set{r,d,r-l(e),d-l(e)}$, which is when $t \in T'$ at least one of the endpoints of the event touches the endpoint of interval $S$.
We claim $T'$ contains all turning points of $h_S$.
Based on the length of the event and interval, there are three cases illustrated in \cref{fig:hS} where in (i) $l(e) < |S|$; in (ii) $l(e) > |S|$ and in (iii) $l(e)=|S|$.
We can draw $h_S$ for each of these three cases and see $T'$ contains all turning points of $h_S$.
\begin{figure}[H]
    \centering
    \includegraphics{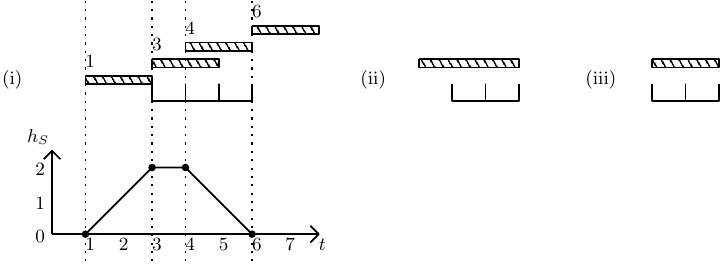}
    \caption{Illustration of three cases in \cref{lem:poly:core-instance:h}. In case (i), $S=[3,6)$ and $l(e)=2$. $T'=\set{3,6,1,4}$.}
    \label{fig:hS}
\end{figure}
\end{proof}

\begin{observation}\label{sumtp}
    Given a set of functions $\cup_{i\in[\kappa]} \set{h_i}$, each $h_i$ is piecewise linear with turning point set $H_i$. Then the turning point set of function $\sum_{i\in[\kappa]} h_i$ is $H=\cup_{i\in[\kappa]} H_i$. Additionally, there must exist a point $t \in H$ such that $\sum_{i\in[\kappa]} h_i$ is maximized.
\end{observation}

\begin{proof}
We sort all points in $H$ in non-decreasing order.
For every two adjacent points in $H$, functions $h_i$ for $i\in[\kappa]$ are all linear and have no turning points and so does $\sum_{i\in[\kappa]} h_i$.
Hence, the turning points of $\sum_{i\in[\kappa]} h_i$ are all contained in $H$.
Since each $h_i$ is piecewise linear and thus so does $\sum_{i\in[\kappa]} h_i$. Therefore, there must exist a point $t \in H$ such that $\sum_{i\in[\kappa]} h_i$ is maximized.
\end{proof}

\begin{lemma}\label{lem:poly:core-instance:agrfig}
The function $g:T\to\N_{\geq 0}$ can be drawn in polynomial time for the core instance $\hcI$.
\end{lemma}

\begin{proof}
As is typical, we show that all turning points of $g$ can be computed in polynomial time.
By \cref{lem:poly:core-instance:h}, we have the turning point set of function $h(I^*_i,\cdot)$ for each $i\in[\kappa]$.
Note that the size of each turning point set is at most $4$ by \cref{lem:poly:core-instance:h}.
Let $H$ be the union of all turning points for these functions and $H$ has size at most $4\kappa$.
By \cref{sumtp}, $H$ is turning point set of function $\sum_{i\in[\kappa]}h(I_{i}^*,t))$.
And since $p^*$ and $|\bigcup_{i\in[\kappa]}I^*_i|$ are constant, the turning points set of function $p^*-(|\bigcup_{i\in[\kappa]}I^*_i|-\sum_{i\in[\kappa]}h(I_{i}^*,t))$ is also $H$ with $|H| \le 4\kappa$.
Due to the same reason, the turning point set of the function $\sum_{j'\in J'}h(I_{j'},\cdot)$ is at most $4\abs{J'}$.
As for the turning points of function $\max\{p^*-(|\bigcup_{i\in[\kappa]}I^*_i|-\sum_{i\in[\kappa]}h(I_{i}^*,t)),0\}$, it will be the union of $H$ and the set $X$ of intersection points between function $p^*-(|\bigcup_{i\in[\kappa]}I^*_i|-\sum_{i\in[\kappa]}h(I_{i}^*,t))$ and $h=0$. 
For every two adjacent points in $H$, there is at most one intersection point between function $p^*-(|\bigcup_{i\in[\kappa]}I^*_i|-\sum_{i\in[\kappa]}h(I_{i}^*,t))$ and $h=0$ and thus the size of $X$ is at most $|H| \le 4\kappa$.
In sum, the size of all turning points in the function $g$ is $\poly(\abs{J'},\kappa)$, and we can also compute them in polynomial time.

\end{proof}

\subsection{General One-event Instance}\label{sec:gen}
To implement the idea of breaking down the general one-event scheduling into subproblems with the core instance $\hcI$, we first see that by \cref{sumtp} if we can know all the turning points of maximum agreement function $g(t)$ of each agent, the optimal event schedule $t$ which maximizes the summation of $g(t)$ can be found by enumerating all the turning points of $g(t)$ of all agents since each $g(t)$ is piecewise linear. 
Therefore, we aim at finding the turning points of each agent's $g(t)$ for $t \in T$ (plotting $g(t)$ for $t \in T$).
To this end, we break down the time span $T$ into polynomial number of disjoint time segments such that when $t$ is in each segment, the general one-event instance with one agent is actually a core instance $\hcI$ (\cref{onevary}) for which we can plot $g(t)$ in polynomial time due to \cref{lem:poly:core-instance:agrfig}.
Hereafter we focus on how to plot one agent's $g(t)$ and suppose this agent has a job set $J$ without using the subscript $i$ to denote the identity of this agent. 
Let $\sch(J,T)$ represent a job schedule of $J$ within timeline $T$ and $\slot(\sch^t(J,T))$ be the set of time slots where jobs in $J$ are processed in schedule $\sch(J,T)$.

\begin{figure}[H]
    \centering
    \includegraphics[width=15cm]{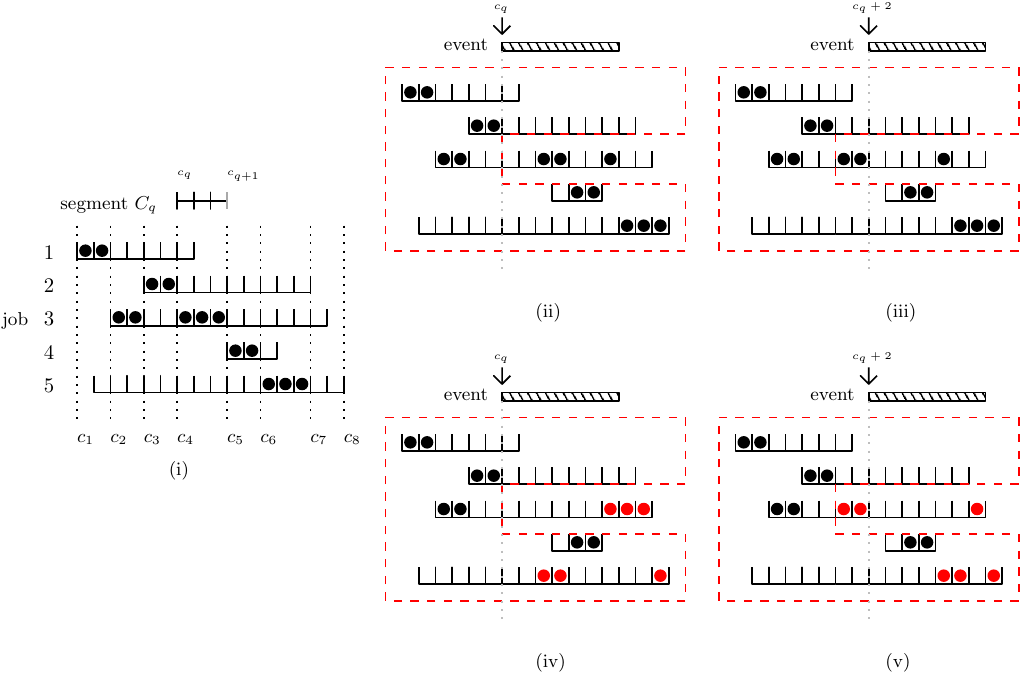}
    \caption{There is an agent's job set $J=\set{1,2,3,4,5}$. An EDF job ordering $\pi$ gives $\pi(1)=1,\pi(4)=2,\pi(2)=3,\pi(3)=4,\pi(1)=5$; and an EDF job ordering in reverse direction gives $\Bar{\pi}(4)=1,\Bar{\pi}(2)=2,\Bar{\pi}(3)=3,\Bar{\pi}(5)=4,\Bar{\pi}(1)=5$. (i) defines a set of time segments based on an EDF schedule $\sch^E$ w.r.t. $\pi$ for $J$. (ii) and (iii) present two optimal job schedules $\sch^{c_q}$ and $\sch^{c_q+2}$ computed by \cref{alg:jobsgevent} with $\pi$, $\Bar{\pi}$ and $\hj=3$ when the event is scheduled at $c_q$ and $c_q+2$. We can see the partial job schedules in the dashed line area are the same in (ii) and (iii). Compared with (ii) and (iii), (iv) and (v) show if there is no priority modification of job $3$ in line \ref{line:biedf_modify_priority} of  \cref{alg:jobsgevent}, job $5$'s schedule will be disturbed by the change of job $3$'s schedule due to the movement of event schedule.}
    \label{fig:biedf}
\end{figure}

\paragraph{Partition time span $T$:} We will do the partition based on an early deadline first (EDF) schedule of $J$. We call $\pi: J \rightarrow \mathbb{N}_+$ an EDF ordering of $J$ if $d_{j_1} \le d_{j_2}$ for any $\pi(j_1) \le \pi(j_2)$. 
An EDF schedule of $J$ with respect to (w.r.t.) $\pi$ can be constructed as follows: go through all time slots in $T$ from the first one to the last one. For $t$-th time slot, among all available jobs $J^a$ whose job intervals contain $t$-th time slot and which haven't been finished, schedule job $j \in J^a$ with the smallest $\pi(j)$ value. 
Suppose $\sch^E$ is such an EDF schedule of $J$.  
$\sch^E$ will partition time span $T$ into job-specific time segments $\cC=\set{C_1,\cdots,C_Q}$ where $C_q=[c_q,c_{q+1})$ , $T=\cup_{q=1}^Q C_q$. There is at most one job scheduled within a segment $C_q$, and jobs scheduled within adjacent time segments are different. See subfigure (i) of \cref{fig:biedf} for an example. Since by EDF, each job can only be interrupted at most $|J|$ times, therefore $Q=O(|J|)$. Then we have the following key lemma:

\begin{lemma}\label{onevary}
    Given an EDF ordering $\pi$ of $J$ and the partition $\cC=\cup_{q=1}^Q$ of timeline based on the EDF schedule $\sch^E$ w.r.t. $\pi$, when $t$ varies in an arbitrary segment $C_q=[c_q,c_{q+1})$, i.e., $t=c_q,\cdots,c_{q+1}-1$, there exist a set of job schedules $\cJ=\cup_{t \in [c_q,c_{q+1})} \set{\sch^t}$ where each $\sch^t$ is an optimal job schedule when the event is scheduled at $t$ and $\sch^t$'s only differ at most one job's schedule. Each $\sch^t$ can be found in polynomial time.
\end{lemma}

\paragraph{Identify core instance $\hcI$:} Within each segment $[c_q,c_{q+1})$, suppose job $\hj$ is the only one job on whose schedule $\sch^t$'s differ and let $J_{-\hj}$ be the set of jobs other than job $\hj$ whose schedules are unchanged in $\cJ$, as described in \cref{onevary} (if $\sch^E$ doesn't schedule any job within $C_q$, let $\hj=\perp$ and $J_{-\perp}=J$). 
For the purpose of plotting $g(t)$, we only care about the optimal job schedule at each $t$ and how it changes at different $t$.
Therefore, for each job $j\in J_{-\hj}$, we can replace it with a rigid job $j'$ occupying time slots in $\slot(\sch^t(j,T))$ which are the time slots that $j$ is scheduled in $\sch^t$ and let $J'$ be the set of all such new rigid jobs;
for job $\hj$, we can replace it with a multi-interval job $j^*$ occupying time slots in $I_{\hj} \backslash \slot(\sch^t(J_{-\hj},T))$ where $I_{\hj}$ is job $\hj$'s interval (if $\hj=\perp$, we do nothing).
Note that $j^*$ and $J'$'s job intervals are non-overlap, and job intervals of each job $j\in J'$ are also non-overlap since they are derived from one agent's feasible job schedule.
Let $\hat{J}=\set{j^*} \cup J'$.
Suppose the original instance is $\cI=(e,J)$ and the new instance is $\cI'=(e,\hat{J})$.
By \cref{onevary}, instance $\cI'$ can be constructed within polynomial time since $\sch^t$ can be constructed in polynomial time (Actually, we only need to compute one $\sch^{c_q}$ for the purpose of constructing $\cI'$).
Instances $\cI$ and $\cI'$ are equivalent in the sense that at each $t$, their $g(t)$ are equal, and thus the figures of their $g(t)$ are the same.
Note that $\cI'$ is a core instance $\hcI$ and thus by \cref{lem:poly:core-instance:agrfig} we can draw $g(t)$ for $t \in [c_q,c_{q+1})$ of $\cI'$ within polynomial time. 
Therefore by ``gluing '' $g(t)$ for $t$ in each segment, we have the following lemma:
\begin{lemma}
    An agent's $g(t)$ for $t \in T$ can be drawn in polynomial time for the general instance.
\end{lemma}

Now we focus on how to prove \cref{onevary}. The core of the proof is finding a set of desired almost-unchanged optimal job schedules $\cJ$ as described in \cref{onevary} within polynomial time. 
However, an event schedule can have many different optimal job schedules, and two adjacent event schedules (such as $[t,t+l(e))$ and $[t+1,t+1+l(e))$) can also have quite different optimal job schedules. 
To bypass this obstacle, below we specially design an algorithm for systematically finding an optimal job schedule given one event being scheduled at each $t$. 
By using this algorithm to find an optimal job schedule for each $t$, we show we can get a set of desirable optimal job schedules $\cJ$ in \cref{onevary}. 
Intuitively, \cref{alg:joboneevent} combines two ``connected'' EDF schedules, one is found in forward ($t=1 \rightarrow t=|T|$) direction in Stage 1, and the other is found in the reverse direction in Stage 2 with jobs modified based on stage 1. 
It is actually trying to greedily schedule jobs' processing time outside of event-occupied time slots and meanwhile to guarantee the feasibility of the schedule \footnote{For the convenience of proof later, we present \cref{alg:joboneevent} as a time slot traverse algorithm. However, \cref{alg:joboneevent} is actually an EDF algorithm with preemption and thus can be implemented within $O(|J|\log |J|)$ time using two priority queues (See Chapter 3 in \cite{DBLP:journals/corr/abs-2001-06005}).}.   
Note that an agent can have different EDF job schedules due to tie breaking on the jobs with the same deadlines. 
Hence we will input arbitrary but specific EDF orderings into the algorithm to guarantee a consistent tie breaking rule in order to minimize the change of solutions when applying \cref{alg:jobsgevent} on different $t$. 
See \cref{fig:biedf} (ii) and (iii) for examples.

\begin{algorithm}[H]
\caption{Optimal job schedule given one event schedule job($(e,t^e),J,T,\pi,\Bar{\pi},\hat{j}$)}\label{alg:joboneevent}
\begin{algorithmic}[1]
\Require Event schedule $(e,t^e)$, one agent's job set $J$, timeline $T$, two orderings $\pi$ and $\Bar{\pi}$ where $\pi$ is according to the ascending order of $J$'s deadlines and $\Bar{\pi}$ is according to the descending order of $J$'s release times, $\hat{j}$ is a job in $J$ satisfying that $\hat{j}$'s job interval $I_{\hj}$ contains time slot $[t^e,t^e+1)$.
\Ensure A job schedule $\sch(J,T)$ for job set $J$.
\For{$t=1:t^e-1$} \Comment{Stage 1}
\State $\cC \leftarrow \set{j \in J: t \in I_j, p_j > 0}$
\State $j^*=\argmin_{j \in \cC} \pi(j)$
\State Schedule $j^*$ at $t$-th time slot in $\sch(J,T)$.
\State $p_{j^*} = p_{j^*} - 1$ 
\EndFor

% \State $r_j = \max \set{t^e,r_j}$ for $j$ with $t^e \in I_j$ \Comment{Modify release time.}

\For{$t=|T|:t^e$} \Comment{Stage 2}
\State $\cC \leftarrow \set{j \in J: t \in I_j, p_j > 0}$
\State Move $\hat{j}$ to the last position in $\Bar{\pi}$ and keep the order of the remain part in $\Bar{\pi}$ unchanged. 
\label{line:biedf_modify_priority}
\State $j^*=\argmin_{j \in \cC} \Bar{\pi}(j)$
\State Schedule $j^*$ at $t$-th time slot in $\sch(J,T)$.
\State $p_{j^*} = p_{j^*} - 1$ 
\EndFor
\end{algorithmic}
\label{alg:jobsgevent}
\end{algorithm}

\begin{lemma}\label{biEDF}
\cref{alg:jobsgevent} finds an optimal job schedule given one event is scheduled at $t^e$.
\end{lemma}

\begin{proof}
    We use an exchange argument to show we can transform an arbitrary optimal job schedule $\sch^o$ into a job schedule $\sch^g$ found by \cref{alg:joboneevent}.

    We compare $\sch^o$ and $\sch^g$ time slot by time slot according to the time slot sequence $A=(1,\cdots,t^e-1,|T|,|T|-1,\cdots,t^e)$ and we say time slot $t_1$ is before (after) time slot $t_2$ if $t_1$ is before (after) $t_2$ in sequence $A$.
    Suppose $t'$-th time slot is the first time slot where the two schedules differ. 
    \paragraph{Case \Rom{1}}: $t' \in [1,t^e-1]$, i.e., $t'$ is the time slot checked in Stage $1$. We further consider three subcases: 
    \begin{itemize}
        \item Subcase 1.1: $\sch^o$ and $\sch^g$ schedule different jobs at $t'$-th time slot.
        Suppose $\sch^o$ schedules job $j_1$ and $\sch^g$ schedules job $j_2$ at $t'$-th time slot. 
        We do a swap for the optimal job schedule $\sch^o$: we find the time slot after $t'$  where $j_2$ is scheduled and schedule $j_1$ there, and then schedule $j_2$ at $t'$. 
        This is feasible since $d_{j_2} \le d_{j_1}$ and thus we can delay $j_1$ and move $j_2$ earlier. Such swap won't change $\slot(\sch^o(J,T))$ and thus won't affect optimality.

        \item Subcase 1.2: $\sch^g$ schedules a job $j$ at $t'$ while $\sch^o$ does not schedule any one job here. 
        Then in $\sch^o$ we find the time slot after $t'$ where job $j$ is scheduled and make it idle. Then we schedule $j$ at $t'$ in $\sch^o$. This can be done because $\sch^g$ schedules job $j$ at $t'$, which indicates $j$ can be scheduled at $t'$. 
        Such modification can only increase agreement since we move a job's processing time from elsewhere to a time slot not occupied by the event schedule. 

        \item Subcase 1.3: $\sch^o$ schedules a job $j$ at $t'$ while $\sch^g$ does not schedule any job here. We claim this situation won't happen due to the greedy rule of \cref{alg:jobsgevent}. 
        Since $j$ is scheduled at $t'$ by $\sch^o$ and before $t'$ $\sch^o$ and $\sch^g$ are totally the same, $j$ will be in the job candidate set $\cC$ for $t'$-th time slot (See the second line of \cref{alg:jobsgevent}) and \cref{alg:jobsgevent} will schedule a job at $t'$, which is a contradiction.
    \end{itemize}

    \paragraph{Case \Rom{2}}: 
    % \lijun{is this ok? feel difficult to describe} 
    $t' \in [t^e,|T|]$, i.e., $t'$ is the time slot checked in Stage $2$. We also further consider the following three subcases. 
    Note that Case \Rom{1} is basically parallel to Case \Rom{2} except for the modification of $\hat{j}$'s priority, and thus, we will mainly show such modification won't affect the transformation of $\sch^o$ to $\sch^g$. 
    The high-level intuition of the feasibility of such priority modification is that in the perspective of reverse direction, all jobs whose intervals contain $t^e$-th time slot have a common actual deadline which is $t^e$ and their deadlines are the latest among all jobs. 
    Hence the decreasing of the priority of $\hj$ is actually just a tie-breaking operation that won't affect the feasibility and optimality of the solution found by \cref{alg:jobsgevent}. 
    Below we give a formal proof:
    \begin{itemize}
        \item Subcase 2.1: $\sch^o$ and $\sch^g$ schedule different jobs at $t'$-th time slot, suppose $\sch^o$ schedules $j_1$ and $\sch^g$ schedules $j_2$. 
        If neither of $j_1$ and $j_2$ is $\hat{j}$, then according to $\Bar{\pi}$, $r_{j_1} \ge r_{j_2}$ and thus we can do a swap for $\sch^o$ as in Subcase 1.1.
        If $j_1 = \hat{j}$ and $r_{j_2} < r_{j_1}$, note that $r_{j_1}=r_{\hat{j}} \le t^e$ and thus $r_{j_2} < r_{j_1} \le t^e$, since $J$ is assumed to have a feasible schedule and again before $t'$ $\sch^o$ and $\sch^g$ are totally the same, then we can also find a time slot after $t'$,i.e., within interval $[t^e,t'-1]$, where $j_2$ is scheduled and schedule $j_1$ there, and then schedule $j_2$ at $t'$. 
        The optimality won't be affected due to the same reason in Subcase 1.1. 

        \item Subcase 2.2: $\sch^g$ schedules a job $j$ at $t'$ while $\sch^o$ does not schedule any one job here. 
        Then in $\sch^o$ we find the time slot after $t'$ where job $j$ is scheduled and make it idle. Then we schedule $j$ at $t'$ in $\sch^o$. Whether $j$ is $\hat{j}$ or not won't affect the feasibility of such modification, and agreement can only increase due to the same reason with subcase 1.2.

        \item Subcase 2.3: Same with Subcase 1.3
    \end{itemize}

    After checking all time slots in the order $A$ we will transform $\sch^o$ into $\sch^g$.
\end{proof}

Now we use \cref{alg:jobsgevent} to prove \cref{onevary}.

\begin{proof}[Proof of \cref{onevary}]
    Let $\hat{j} \in J \cup \set{\perp}$ be the job scheduled within segment $[c_q,c_{q+1})$ in the EDF schedule $\sch^E$.
    Let $\cJ=\cup_{t \in [c_q,c_{q+1})} \set{\sch^t}$ where each $\sch^t$ is found by \cref{alg:jobsgevent} with input $((e,t),J,T,\pi,\Bar{\pi},\hat{j})$ where only $t$ of the input varies and $\pi$ is required to be the ordering which produces $\sch^E$.
    \cref{alg:joboneevent} is actually an EDF algorithm with preemption which is known to have polynomial implementation using two priority queues and thus each $\sch^t$ can be computed in polynomial time. 
    Since within time area $[1,t)$, $\sch^E$ and $\sch^t$ are produced by applying the same job ordering $\pi$ on the same job set, we have the following simple observation:
    
    \begin{observation}\label{edfalign}
        $\sch^E$ and $\sch^t$ are always the same within time area $[1,t)$.
    \end{observation}
    
    We then compare $\sch^{c_q}$ and an arbitrary $\sch^t$ for $t \in (c_q,c_{q+1})$ and show their difference only lies in job $\hat{j}$'s schedule (if $\hj=\perp$, $\sch^{c_q}$ and $\sch^t$ are completely the same).
    We compare $\sch^{c_q}$ and $\sch^t$ in three time areas:
    \begin{enumerate}
        \item $[1,c_q)$: It's easy to see $\sch^{c_q}$ and $\sch^t$ have the same schedules within time area $[1,c_q)$ since both $\sch^{c_q}$ and $\sch^t$ are the same with $\sch^E$ within $[1,c_q)$ by \cref{edfalign}.
        \item $[c_q,t)$: We claim the difference of $\sch^{c_q}$ and $\sch^t$ within time area $[c_q,t)$ only lies in job $\hat{j}$'s schedule.
        First by \cref{edfalign}, in $\sch^t$, job $\hat{j}$ must be scheduled within $[c_q,t)$ since $\hat{j}$ is scheduled within $[c_q,t)$ in $\sch^E$.
        In $\sch^{c_q}$, suppose for contradiction there exist some jobs not $\hat{j}$ scheduled at time slots in $[c_q,t)$ and let $j'$ be one of such jobs scheduled at $t' \in [c_q,t)$.
        When \cref{alg:jobsgevent} reaches the step $t=t'$ in Line $7$, since $t' \in I_{\hat{j}}$ and $\hat{j}$ has the lowest priority among all jobs, the fact that \cref{alg:jobsgevent} schedules $j'$ at $t$ indicates $\hat{j}$ has been finished when \cref{alg:jobsgevent} reaches $t=t'$.
        According to $\sch^E$ we know in $\sch^{c_q}$ $\hat{j}$ must occupy at least $t-c_q$ time slots within $[c_q,|T|]$. 
        Since now $j'$ occupies a time slot within $[c_q,t-1]$, there must exist a time slot $t'' \in [t,|T|]$ where $\hat{j}$ is scheduled. 
        We have $j'$'s deadline must be at least larger than $t''$; otherwise, $j'$'s deadline will be smaller than $\hat{j}$'s, contradicting the fact that in $\sch^E$ $\hat{j}$ is scheduled at $t'$ when both $\hat{j}$ and $j'$ are in the candidate job set. 
        Therefore, when \cref{alg:jobsgevent} reaches the step $t=t''$, both $j'$ and $\hat{j}$ are in the candidate job set, and \cref{alg:jobsgevent} schedules $\hat{j}$ instead of $j'$, contradicting with $\hat{j}$ having the lowest priority.
        Then we know it's impossible that there are jobs other than $\hat{j}$ scheduled within $[c_q,t-1]$ in $\sch^{c_q}$.
        \item $[t,|T|]$: Here the priority modification of job $\hj$ plays an important role since by this operation, the change of job $\hj$'s processing time left for time area $[t,|T|]$ won't affect other jobs' schedule. 
        
        For job set $J_{-\hat{j}}=J \backslash \set{\hat{j}}$, no matter the first for loop of \cref{alg:jobsgevent} goes through $1:c_q-1$ or $1:t-1$, the processing time of $J_{-\hat{j}}$ which are left unfinished after the first loop are the same, since within $[c_q,t)$ only job $\hat{j}$ is scheduled.
        Therefore, when \cref{alg:jobsgevent} checks any one time slot within $[t,|T|]$ the candidate job set contains the same jobs in $J_{-\hat{j}}$ when $t^e=c_q$ and $t^e=t$.
        Since $\hat{j}$ has the lowest priority, if the candidate job set contains a subset $J'$ of jobs in $J_{-\hat{j}}$, no matter $\hat{j}$ is also contained in the candidate job set or not, \cref{alg:jobsgevent} will select one job from $j'$ to schedule.
    \end{enumerate}
    
\end{proof}

\section{Conclusion}

In this paper, we study the problem of public event scheduling with busy agents, in which a set of public events are required to be scheduled to let agents attend for a period of time as long as possible. We present a general algorithmic framework and give a polynomial time algorithm with a constant approximation ratio for such an NP-hard problem.

This work points out many interesting future directions.
Firstly, it would be interesting to see a better approximation algorithm or a lower bound.
Secondly, our $\frac{1}{2}$-approximate algorithm for general cases uses an involved optimal algorithm for one-event instances.
So, a simplified algorithm for one-event instances would be beneficial, even if some approximation ratio needs to be sacrificed.
Lastly, the weighted setting of our problem is also interesting, i.e., different agents hold a distinct preference for the same event.

\paragraph{Acknowledgement.}
We thank the anonymous reviewers for their many insightful suggestions.
Bo Li is funded by HKSAR RGC (No. PolyU 15224823) and GDSTC (No. 2023A1515010592).
Ruilong Zhang was supported by NSF grant CCF-1844890.

\newpage
\clearpage
\printbibliography

\newpage
\clearpage
\appendix

\end{document}